\documentclass[twocolumn]{aastex63}
\usepackage{float, graphicx, subfigure, booktabs}
\usepackage{amsmath, supertabular, tablefootnote}
\usepackage{threeparttablex}

\shorttitle{FAST Observations of FRB~20220912A}
\shortauthors{Zhang et al.}

\begin{document}
\title{FAST Observations of FRB~20220912A: Burst Properties and Polarization Characteristics}

\correspondingauthor{Yongkun Zhang, Di Li, Bing Zhang} \email{ykzhang@nao.cas.cn, dili@nao.cas.cn, bing.zhang@unlv.edu}

\author{Yong-Kun Zhang}
\affil{National Astronomical Observatories, Chinese Academy of Sciences, Beijing 100101, China}
\affil{University of Chinese Academy of Sciences, Beijing 100049, China}
\author{Di Li}
\affil{National Astronomical Observatories, Chinese Academy of Sciences, Beijing 100101, China}
\affil{University of Chinese Academy of Sciences, Beijing 100049, China}
\affil{Research Center for Intelligent Computing Platforms, Zhejiang Laboratory, Hangzhou 311100, China}
\affil{NAOC-UKZN Computational Astrophysics Centre, University of KwaZulu-Natal, Durban 4000, South Africa}
\author{Bing Zhang}
\affil{Nevada Center for Astrophysics, University of Nevada, Las Vegas, NV 89154, USA}
\affil{Department of Physics and Astronomy, University of Nevada, Las Vegas, NV 89154, USA}
\author{Shuo Cao}
\affil{Yunnan Observatories, Chinese Academy of Sciences, Kunming 650216, China}
\author{Yi Feng}
\affil{Research Center for Intelligent Computing Platforms, Zhejiang Laboratory, Hangzhou 311100, China}
\author{Wei-Yang Wang}
\affil{University of Chinese Academy of Sciences, Beijing 100049, China}
\affil{Department of Astronomy, Peking University, Beijing 100871, China}
\author{Yuanhong Qu}
\affil{Nevada Center for Astrophysics, University of Nevada, Las Vegas, NV 89154, USA}
\affil{Department of Physics and Astronomy, University of Nevada, Las Vegas, NV 89154, USA}
\author{Jia-Rui Niu}
\affil{National Astronomical Observatories, Chinese Academy of Sciences, Beijing 100101, China}
\affil{University of Chinese Academy of Sciences, Beijing 100049, China}
\author{Wei-Wei Zhu}
\affil{National Astronomical Observatories, Chinese Academy of Sciences, Beijing 100101, China}
\author{Jin-Lin Han}
\affil{National Astronomical Observatories, Chinese Academy of Sciences, Beijing 100101, China}
\author{Peng Jiang}
\affil{National Astronomical Observatories, Chinese Academy of Sciences, Beijing 100101, China}
\affil{CAS Key Laboratory of FAST, National Astronomical Observatories, Chinese Academy of Sciences, Beijing 100101, China}
\author{Ke-Jia Lee}
\affil{National Astronomical Observatories, Chinese Academy of Sciences, Beijing 100101, China}
\affil{Department of Astronomy, Peking University, Beijing 100871, China}
\author{Dong-Zi Li}
\affil{Cahill Center for Astronomy and Astrophysics, MC 249-17, California Institute of Technology, Pasadena, CA 91125, USA}
\author{Rui Luo}
\affil{CSIRO Space and Astronomy, PO Box 76, Epping, NSW 1710, Australia}
\affil{Department of Astronomy, School of Physics and Materials Science, Guangzhou University, Guangzhou 510006, China}
\author{Chen-Hui Niu}
\affil{National Astronomical Observatories, Chinese Academy of Sciences, Beijing 100101, China}
\author{Chao-Wei Tsai}
\affil{National Astronomical Observatories, Chinese Academy of Sciences, Beijing 100101, China}
\author{Pei Wang}
\affil{National Astronomical Observatories, Chinese Academy of Sciences, Beijing 100101, China}
\author{Fa-Yin Wang}
\affil{School of Astronomy and Space Science, Nanjing University, Nanjing 210093, China}
\affil{Key Laboratory of Modern Astronomy and Astrophysics (Nanjing University), Ministry of Education, China}
\author{Zi-Wei Wu}
\affil{National Astronomical Observatories, Chinese Academy of Sciences, Beijing 100101, China}
\author{Heng Xu}
\affil{National Astronomical Observatories, Chinese Academy of Sciences, Beijing 100101, China}
\author{Yuan-Pei Yang}
\affil{South-Western Institute for Astronomy Research, Yunnan University, Kunming, Yunnan 650504, China}
\affil{Purple Mountain Observatory, Chinese Academy of Sciences, Nanjing, Jiangsu 210023, China}
\author{Jun-Shuo Zhang}
\affil{National Astronomical Observatories, Chinese Academy of Sciences, Beijing 100101, China}
\affil{University of Chinese Academy of Sciences, Beijing 100049, China}
\author{De-Jiang Zhou}
\affil{National Astronomical Observatories, Chinese Academy of Sciences, Beijing 100101, China}
\affil{University of Chinese Academy of Sciences, Beijing 100049, China}
\author{Yu-Hao Zhu}
\affil{National Astronomical Observatories, Chinese Academy of Sciences, Beijing 100101, China}
\affil{University of Chinese Academy of Sciences, Beijing 100049, China}

\begin{abstract}
We report the observations of FRB~20220912A using the Five-hundred-meter Aperture Spherical radio Telescope (FAST). We conducted 17 observations totaling 8.67 hours and detected a total of 1076 bursts with an event rate up to 390 hr$^{-1}$. The cumulative energy distribution can be well described using a broken power-law function with the lower and higher-energy slopes of $-0.38\pm0.02$ and $-2.07\pm0.07$, respectively. We also report the L band ($1-1.5$ GHz) spectral index of the synthetic spectrum of FRB~20220912A bursts, which is $-2.6\pm0.21$. The average rotation measure (RM) value of the bursts from FRB~20220912A is $-0.08\pm5.39\ \rm rad\,m^{-2}$, close to 0 $\rm rad\,m^{-2}$ and maintain relatively stable over two months. Most bursts have nearly 100\% linear polarization. About 45\% of the bursts have circular polarization with SNR $>$ 3, and the highest circular polarization degree can reach 70\%. Our observations suggest that FRB~20220912A is located in a relatively clean local environment with complex circular polarization characteristics. These various behaviors imply that the mechanism of circular polarization of FRBs likely originates from an intrinsic radiation mechanism, such as coherent curvature radiation or inverse Compton scattering inside the magnetosphere of the FRB engine source (e.g. a magnetar).
\end{abstract}

\section{Introduction} \label{sec:intro}

Fast radio bursts (FRBs) are a type of astronomical phenomenon characterized by brief, intense pulses of radio waves from unknown sources. Since their discovery in 2007 \citep{2007Sci...318..777L}, FRBs have remained a mystery, and their origins are still unknown. 

To gain further understanding of the origins and radiation mechanisms involved, it is crucial to conduct statistical analysis and investigate the properties of a large sample of bursts. FRBs are empirically classified into two categories: non-repeating FRBs and repeating FRBs, with the latter accounting for a small fraction of the entire FRB population, and only a handful of them exhibiting event rates of several tens to hundreds per hour, such as FRB~20121102A \citep{2021Natur.598..267L, 2023MNRAS.519..666J}, FRB~20200120E \citep{2023MNRAS.520.2281N}, and FRB~20201124A \citep{2022Natur.609..685X, 2022RAA....22l4002Z}. The study of FRB polarization may reveal the complexity of the local environment. FRB~20121102A and FRB~20190520B both show exceptionally high and variable RM \citep{2018Natur.553..182M, 2021ApJ...908L..10H, 2022arXiv220308151D, 2022arXiv220211112A}, while FRB~20201124A displays short-time irregular RM oscillations \cite{2022Natur.609..685X}. The RM variation of FRB~20180916B also exceeded 40\% \citep{2022arXiv220509221M}. Recently, CHIME reported measurements of the polarization of 12 repeating FRBs, finding that a significant proportion of FRBs experience RM changes of tens to hundreds within months \citep{2023arXiv230208386M}. These facts suggest that most FRB progenitors may be located in a complex, dynamically evolving magnetized environment, such as a supernova remnant, a pulsar wind nebula, or a binary system with a massive companion star \citep{2022Sci...375.1266F, 2022NatCo..13.4382W,2023ApJ...942..102Z,2023MNRAS.520.2039Y}.

In October 2022, the CHIME/FRB collaboration reported a new FRB, named FRB~20220912A \citep{2022ATel15679....1M}. Over the course of three days, nine bursts were detected in the CHIME band, leading to the expectation that this may become a highly active repeating FRB. According to \cite{2022ATel15679....1M}, the dispersion measure (DM) of FRB~20220912A is $219.46\,\rm pc\, cm^{-3}$, with a small RM value of $0.6\,\rm rad\, m^{-2}$. The high activity level of FRB~20220912A allowed the DSA-110 collaboration to quickly localize the source in a host galaxy with a redshift of 0.077 \citep{2022arXiv221109049R}. It is speculated that this host galaxy contributes less than $50\,\rm pc\, cm^{-3}$ to DM, considering a Galactic DM contribution of $125\,\rm pc\, cm^{-3}$ \citep{2002astro.ph..7156C, 2017ApJ...835...29Y} plus a Milky Way halo contribution of $10\,\rm pc\, cm^{-3}$ \citep{2020MNRAS.496L.106K}. The low host galaxy DM contribution is in contrast to the host galaxies of FRB~20121102A and FRB~20190520B which contribute a significant amount of DM \citep{2017Natur.541...58C, 2022Natur.606..873N}. Since FRB~20220912A's discovery, numerous telescopes and telescope arrays detected bursts from FRB~20220912A \citep{2022ATel15691....1H, 2022ATel15693....1R, 2022ATel15695....1P, 2022ATel15713....1F, 2022ATel15723....1F, 2022ATel15727....1K, 2022ATel15733....1Z, 2022ATel15734....1P, 2022ATel15735....1S, 2022ATel15758....1Y, 2022ATel15791....1R, 2022ATel15806....1B, 2022ATel15817....1O}, attesting to its high brightness and activity.

Here we report on the FAST observation of the active repeating FRB~20220912A. Our observations and data processing procedures are described in Section~\ref{sec:data}. Our results are presented in Section~\ref{sec:result}. We discuss the circular polarization expressions in Section~\ref{sec:disc} and conclude in Section~\ref{sec:conclusion}.

\section{Observations and data processing} \label{sec:data}

FRB 20220912A was observed since October 28th, 2022, using the center beam of the FAST 19 beam receiver \citep{FAST19Beam} pointing to the coordinate of RA=23$^{\rm h}$09$^{\rm m}$04.9$^{\rm s}$, Dec=+48$^\circ$42$'$25.4$''$ reported by DSA-100 \citep{2022arXiv221109049R}. In 2022, 17 observations with a total of 8.67-hour exposure time were carried out. A high-cadence CAL signal was periodically injected during the first minute of observation for the following flux and polarization calibration. The data was recorded in \textsc{fits} format with a time resolution of 49.152\,$\rm \mu s$, covering the frequency bandwidth from 1 to 1.5\,GHz with 4096 frequency channels.

We use the same pipeline in \cite{2022RAA....22l4002Z} to perform offline burst searches. According to the CHIME report, the DM value of FRB~20220912A is 219.6 $\rm pc\, cm^{-3}$. Data were de-dispersed using this DM and then were identified by a binary classification model whether a burst existed in a data segment. A total of 1076 bursts were detected. %

We estimated the flux density of each burst using the radiometer equation with system temperature $T_{\rm sys}$ and telescope gain $G$ modeled as a function of the zenith angle and observation frequency in \cite{2020RAA....20...64J}. The burst profile is the average flux density that goes over the full observation frequency band. The peak flux $S_{\rm peak}$ is the max value of the burst profile. The burst fluence $F$ is computed by integrating the burst profile with respect to time, and the equivalent width is computed by dividing the fluence by the peak flux. The energy is calculated using the equation \footnote{The calculation of an FRB isotropic energy depends on the spectral shape of the burst \citep[][for a discussion]{zhang22}. If the spectrum is wide (power-law like), it is more appropriate to use the central frequency rather than band width to estimate the energy. If the spectrum is narrow, especially with a measurable width within the telescope bandpass, it is more appropriate to use the bandwidth in the calculation. The repeating bursts typically have narrow spectra, so it is more appropriate to use Eq.(\ref{eq:energy}) to perform the calculations.}\label{footnote:spectrum}

\begin{equation}\label{eq:energy}
    E = 10^{39} {\rm erg}\frac{4\pi}{1+z} \left(\frac{D_L}{10^{28}{\rm cm}}\right)^{2} \left(\frac{F}{\rm Jy\cdot ms}\right)\left(\frac{\Delta\nu}{\rm GHz}\right),
\end{equation}
where $F$ is the fluence obtained, and $\Delta \nu=500\,\rm MHz$ is the observation bandwidth. $D_L=360.86\,\rm Mpc$ is the luminosity distance of FRB~20220912A corresponding to the redshift $z=0.0771$ \citep{2022arXiv221109049R} adopting the standard Planck cosmological model \citep{2016A&A...594A..13P}.

Polarization calibration was achieved by correcting for the differential gain and phase between the receptors through separate measurements of a noise diode signal injected at an angle of $45^{\circ}$ from the linear receptors with the single-axis model using the \textsc{PSRCHIVE} software package.

\section{Results} \label{sec:result}
\subsection{Burst rate and time series analysis} \label{sec:rates}

\begin{figure*}[!htp]
    \centering
    \includegraphics[width=0.8\textwidth]{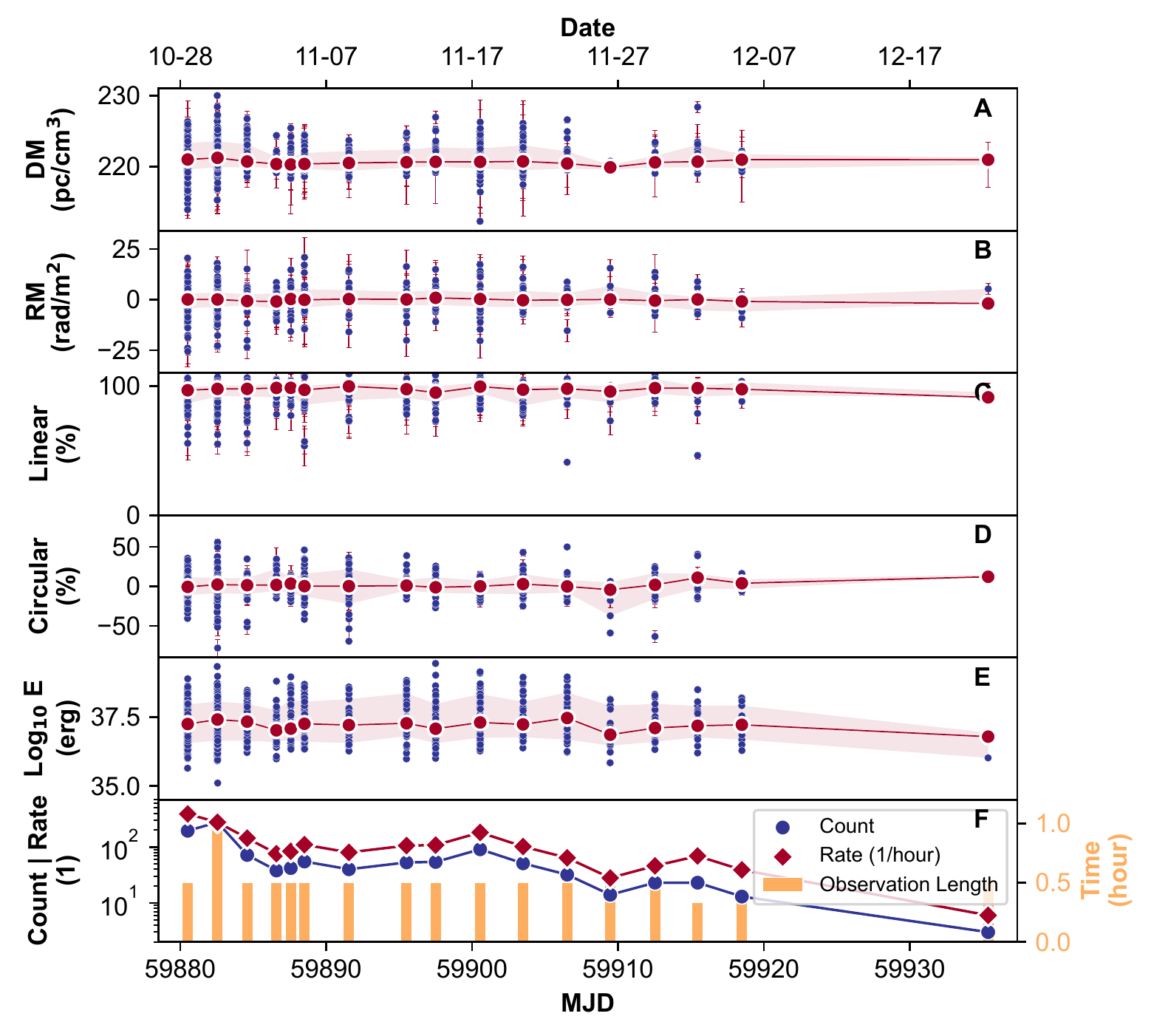}
    \caption{The properties of the bursts from FRB 20220912A observed by FAST. \textbf{A}, \textbf{B}, \textbf{C}, \textbf{D}, and \textbf{E} respectively denote the DM, RM, linear polarization, circular polarization, and energy of the bursts. Each burst is depicted by a blue point with error bars, the red points represent the daily median, and the light red area encompasses the 1$\sigma$ range. \textbf{F} displays the number of bursts detected (blue) and the event rate (red), while the yellow bar symbolizes the observation length.}
    \label{fig:oblog}
\end{figure*}

Figure~\ref{fig:oblog}F displays the length of each observation, along with the detected burst numbers and event rates. Over the course of 17 observations, we detected a total of 1076 bursts, which can be found in Table~\ref{tab:burst}. The event rates of 8 observations exceeded 100 hr$^{-1}$, with the highest event rate of 390 hr$^{-1}$ during the first observation, which is only smaller than FRB20201124A's 542 hr$^{-1}$ \citep{2022RAA....22l4002Z}, demonstrating that FRB~20220912A is a highly active repeating FRB.

\begin{figure}[!htp]
    \centering
    \includegraphics[width=0.45\textwidth]{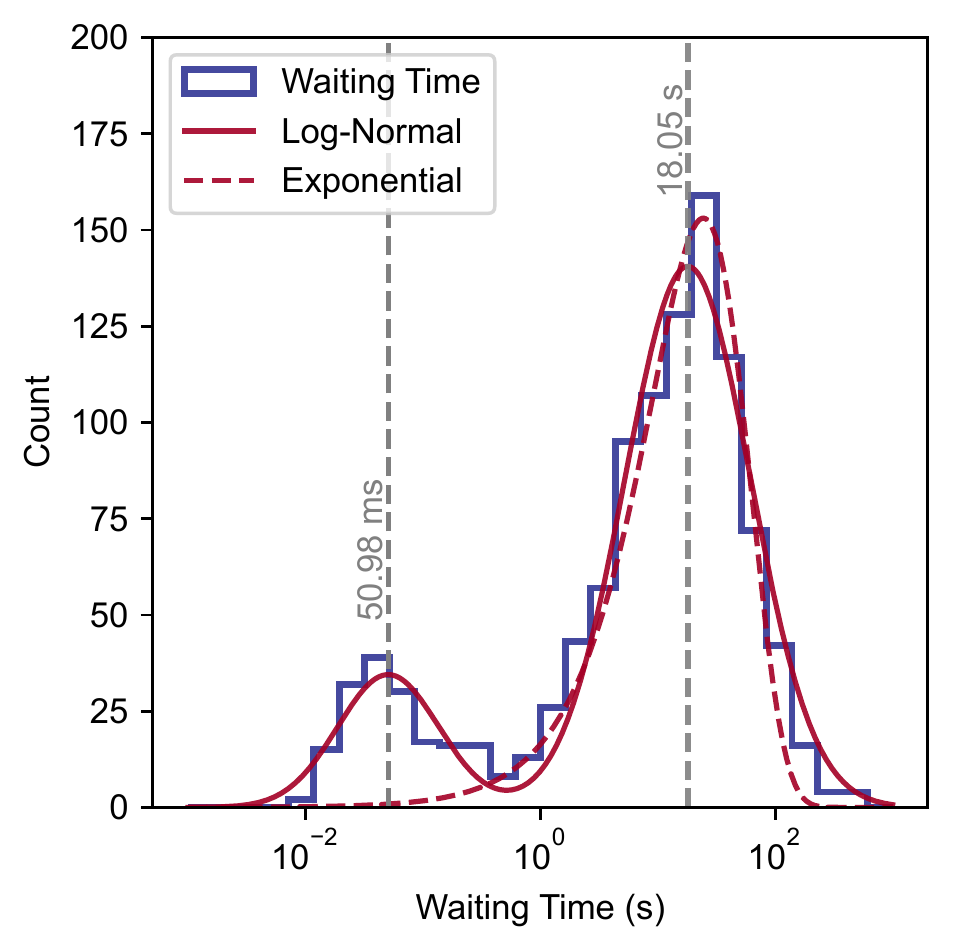}
    \caption{The waiting time distribution of FRB~20220912A. The red solid line representing two Log-Normal functions fitting and the red dashed line indicating an exponential function fitting.}
    \label{fig:wt}
\end{figure}

We calculated the waiting times between bursts for each observation. Similar to FRB~20121102A and FRB~20201124A, FRB~20220912A also exhibits a distinctive bimodal distribution (Figure~\ref{fig:wt}). We utilized two Log-Normal functions to model the waiting time distribution, with peaks located around 18~s and 50~ms, respectively. %

The waiting times of a Poisson process are exponentially distributed. Here, we use an exponential function to fit the waiting time distribution using the waiting times up to longer than the second valley of the two Log-Normal distribution ($\sim$0.52~s). We utilize a Kolmogorov-Smirnov (K-S) test to evaluate the goodness of the Log-Normal and exponential fittings, with p-values of 0.969 and 0.963, respectively, indicating that both models effectively describe the distribution. The Poisson process rate as obtained from the exponential distribution is 0.041 s$^{-1}$ or 147 hr$^{-1}$, which is close to the average observed event rate of 1076/8.67$\sim$124 hr$^{-1}$. The right peak of the waiting time represents the activity of the FRB source during the statistical period. The Log-Normal provides a left peak of the waiting time near 50~ms, which is quite similar to FRB~20201124A (39~ms in \citealt{2022Natur.609..685X} and 51~ms in \citealt{2022RAA....22l4002Z}). The left peak of FRB~20121102A is about 3~ms \citep{2021Natur.598..267L}, significantly different from FRB~20220912A here. However, FRB~20121102A appears to have a weak secondary peak around 50~ms (Figure 3 in \citealt{2021Natur.598..267L}). The characteristic waiting time of 50~ms may signify some fundamental properties of the FRB source emitting the bursts.

Searching for periodicity of FRBs remains an active research area for FRBs. There have been three non-repeating FRBs discovered to have millisecond-level quasi-periods \citep{2022Natur.607..256C}. For repeating FRBs, however, there are no reports of short-period emerged \citep{2018ApJ...866..149Z, 2021Natur.598..267L, 2022Natur.606..873N, 2022Natur.609..685X}. Here, we also conducted a period search for FRB~20220912A using two methods: Lomb-Scargle periodograms (LSP) and phase folding. The LSP method has been widely applied to non-uniformly sampled time series, making it suitable for periodic searches in the arrival time series of FRBs. Phase folding is also a simple, foundational method for period searches. Here, we assumed $\rm MJD-59880$ as the initial phase and computed the phase of each burst under a preset period, counting the longest continuous phase interval without burst existence, i.e., the void fraction. We iterated over periods ranging from 1~ms to 1000~s and computed the void fraction under these periods. A larger void fraction means a more concentrated burst distribution in the phase space, indicating higher reliability of the corresponding period. This approach is similar to the statistics used in \cite{2020MNRAS.495.3551R}, except that we directly calculate the phase of the burst, making it more efficient and accurate. Unfortunately, neither of these two methods yielded a valid periodic signal.

Only FRB~20121102A and FRB~20180916B have been reported to possess potential periodicities, the former approximately 157 days \citep{2020MNRAS.495.3551R, 2021MNRAS.500..448C}, the latter approximately 16 days \citep{2020Natur.582..351C}. These periodicities are composed of active and quiescent phases (like square waves), and not all active phases had bursts detection. Bursts from FRB~20220912A have been detected in all of our observations, precluding exploration of such active-quiescent periods. However, we can define the event rate various over time as a ``light curve'' to search for possible periodicities in activity levels. Given the 54-day duration of FAST observations, we can only search for periodicities up to 27 days. To avoid observational interference on 1-day period, we commence our search from 2-day period. Utilizing Lomb-Scargle periodograms, no reliable periodicities were found within the period range between 2 days and 27 days.

\subsection{Energy} \label{sec:energy}

\begin{figure}[!htp]
    \centering
    \includegraphics[width=0.45\textwidth]{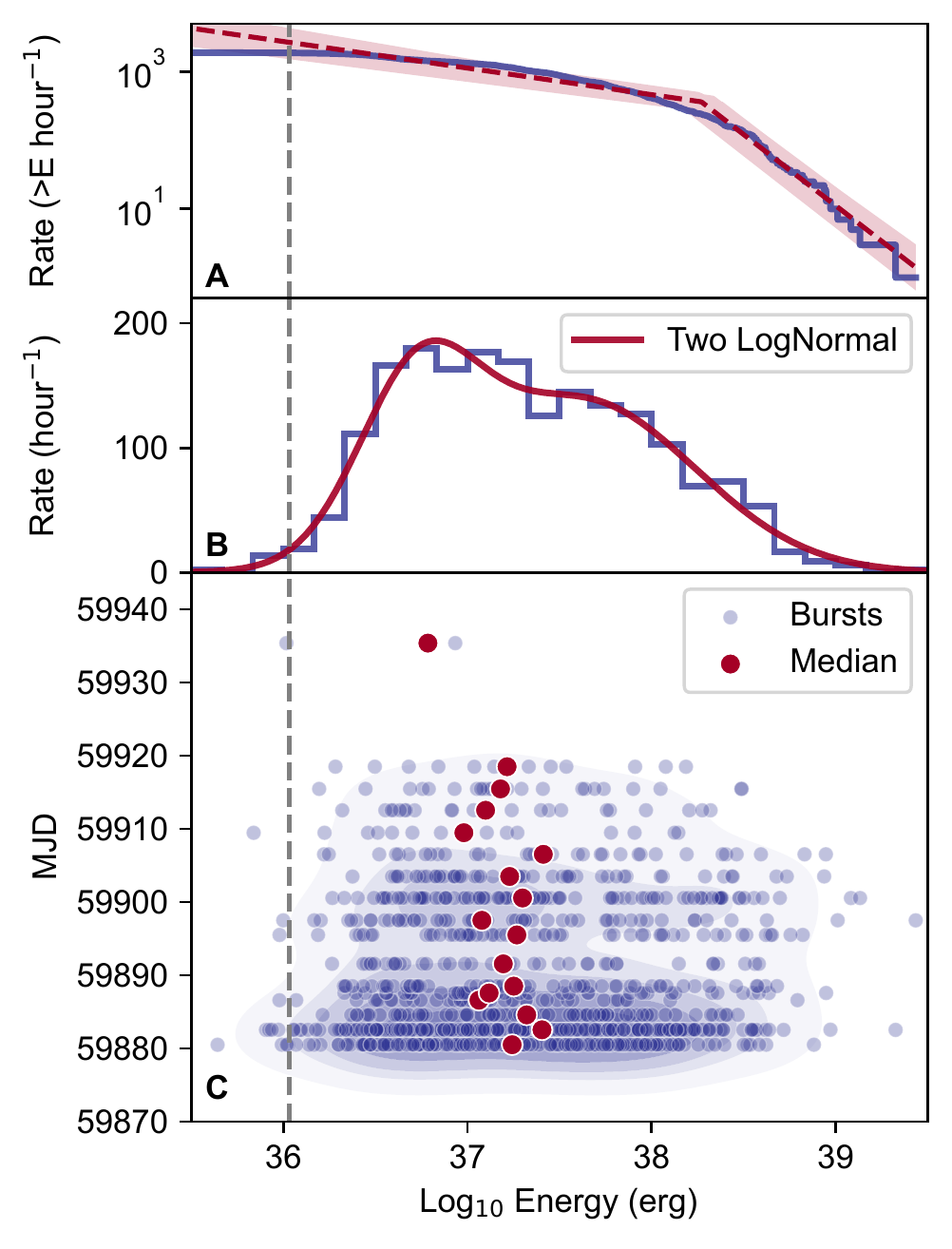}
    \caption{Energy distribution of FRB~20220912A. \textbf{A}: The blue step represents the cumulative probability distribution of the energy function, while the white line denotes the fitting using a broken power-law function. The red region indicates the 1$\sigma$ range of the fitting. \textbf{B}: The differential probability distribution of the energy function, with the red line showing the fitting using two Log-Normal functions. \textbf{C}: The time-dependent burst energy distribution. The blue dots display the energy of 1076 bursts; red dots denote the median energy; and the blue contour depict a 2D kernel density estimation (KDE) of the bursts. The gray dashed line represents the 90\% detection threshold.}
    \label{fig:energy}
\end{figure}

Energy is one of the basic properties of FRBs, which is a physical quantity that can directly reflect the radiation mechanism of FRBs. The energy function of FRBs is typically modeled as a power-law function, probably with a cutoff at the high end \citep[e.g.][]{luo2018,luo2020b,Lu2020,zhangrc2021}. \cite{2021Natur.598..267L}'s detection of the low-energy outburst of FRB~20121102A reveals the multiple radiation mechanisms that FRBs may possess. Figure~\ref{fig:energy} displays the energy function of FRB~20220912A, along with its distribution over time. Due to varying observation lengths, the energy function is weighted based on the observation time. Like FRB~20121102A and FRB~20201124A \citep{2021Natur.598..267L, 2021ApJ...922..115A, 2023MNRAS.519..666J, 2022Natur.609..685X, 2022RAA....22l4002Z}, the differential energy function of FRB~20220912A cannot be explained using one single function. Two Log-Normal functions are used for the fitting, with the corresponding characteristic energies being $5.29\times10^{36}$ erg and $4.13\times10^{37}$ erg, respectively. The integral energy function also cannot be fit with a single power law. Two sections of power law fit are utilized, with the power-law exponents being $-0.38\pm0.02$ and $-2.07\pm0.07$. In Figure~\ref{fig:energy}C, the energy appears to exhibit an evolutionary characteristic over time, with fewer high-energy bursts observed in the latter observations.

The distribution of the center frequency and bandwidth of the FRB~20220912A bursts. For the center frequency distribution, a 3-component Gaussian fit was employed, with peak values of 1.02, 1.08, and 1.38 GHz respectively. As for the distribution of both the center frequency and bandwidth, the bandwidth was fitted by a Gaussian function with 2 sigma. The light yellow background denotes the observed bandwidth of FAST from 1 to 1.5 GHz, while the light blue background limits the effective bandwidth of FAST to 0.5 GHz. Regarding the bandwidth distribution, the LogNormal fitting peaked at 120 MHz. Lastly, the LogNormal fitting of the bandwidth-to-center-frequency ratio had a peak value of 0.11.

After calibrating the time-frequency data, we calculate the fluence of each burst by integrating the flux with respect to time in each frequency channel. Then we average the fluence of all bursts to obtain the synthetic fluence-frequency spectrum of FRB~20220912A. As the different frequency channels of these bursts may be masked as RFI channels, the number of bursts used for averaging fluence varies in frequency channels. We use Poisson counting error as the statistical error and use a simple power-law function to fit the spectral index of the synthetic spectrum.

\begin{equation} \label{eq:spin}
    I = A \times F^{\alpha} + C
\end{equation}

\begin{figure}[!htp]
    \centering
    \includegraphics[width=0.45\textwidth]{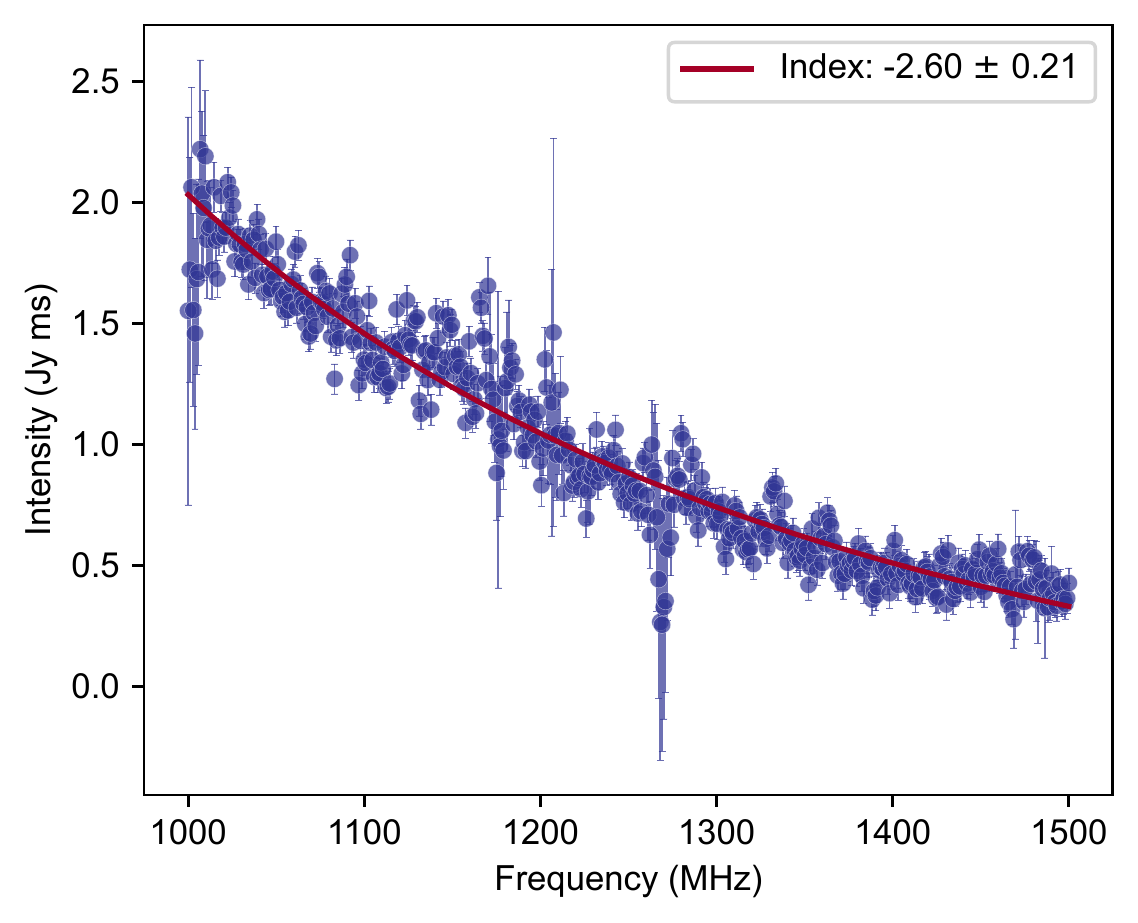}
    \caption{Power-law fitting (red line) to the synthetic spectrum of FRB~20220912A.}
    \label{fig:spectral}
\end{figure}

In Figure~\ref{fig:spectral} one can see large error bars near 1000MHz, 1200MHz, etc. due to frequent RFI interference, resulting in fewer effective data points. For the fit, we randomly select 90\% of the data points from the available data and perform 1000 fittings, resulting in the expectation and error of the FRB~20220912A spectral index being $\alpha=-2.60\pm0.21$. This is the first measurement of the average spectral index of the synthetic spectrum of an FRB source. %
Such a spectrum can be in principle to test against FRB radiation mechanisms (e.g. curvature radiation, \citep{2018ApJ...868...31Y}). 

\begin{figure}[!htp]
    \centering
    \includegraphics[width=0.45\textwidth]{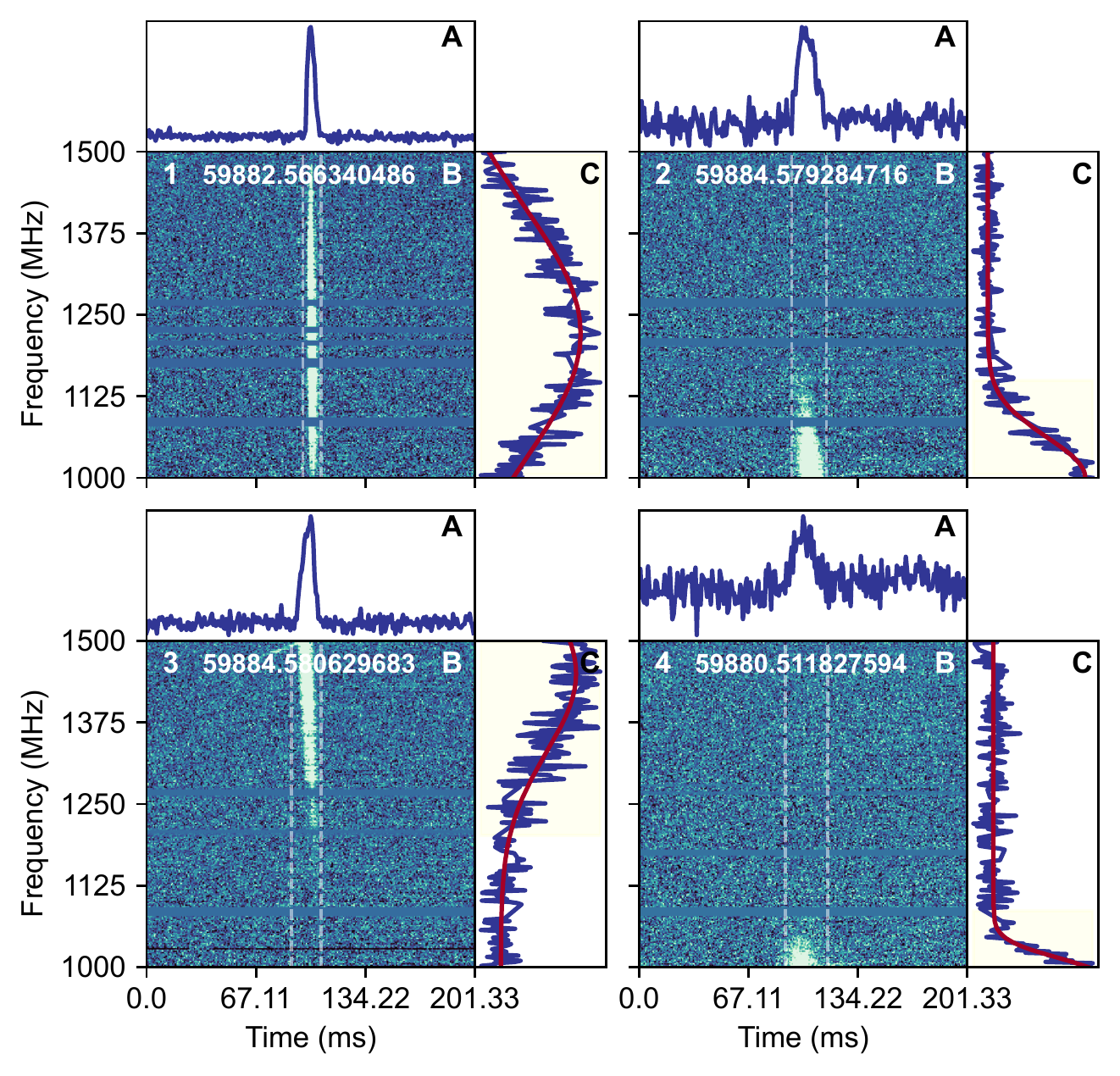} 
    \caption{Bandwidth fitting examples of the FRB~20220912A bursts. \textbf{1}: Burst with FWHM completely falling within the range of 1-1.5 GHz. \textbf{2}: Burst with truncation at 1 GHz, resulting in incomplete recording. \textbf{3}: Burst with truncation at 1.5 GHz, resulting in incomplete recording. \textbf{4}: Burst with truncation at 1 GHz, hindering visibility of Gaussian peak.}
    \label{fig:freqplot}
\end{figure}

Although the synthetic spectrum of all bursts conforms well to a power-law distribution, it is difficult to describe the relatively narrow bandwidth of individual bursts using a power-law model. Based on the current empirical evidence, a Gaussian function may be a viable alternative to describe the  spectra of individual bursts. Prior research has employed a Gaussian model to investigate FRB~20121102A, with the fitting residual demonstrating the potential efficacy of the Gaussian distribution as a model, as illustrated in Figure 5 of \cite{2021ApJ...922..115A}. Similarly, \cite{2022RAA....22l4001Z} fitted the spectrum of FRB~20201124A with Gaussian functions and identified a bimodal distribution of central frequencies. Here, we also attempt to fit the spectrum of individual bursts using a Gaussian function

\begin{equation} \label{eq:gaus}
    I = A \times \exp\left[-\frac{(F-\mu)^2}{2\sigma^2}\right]
\end{equation}
where we use the fitted $\mu$ as the center frequency $\nu_0$, and the full-width-at-half-maximum (FWHM=$2\sqrt{2{\ln}2}\sigma$) as the bandwidth $\Delta \nu$. It should be noted that due to the limited bandwidth of FAST, some bursts likely have emission outside the FAST band and not fully recorded. This introduces additional uncertainties to the bandwidth fitting results. Four examples of bandwidth fitting are illustrated in Figure~\ref{fig:freqplot}.

\begin{figure}[!htp]
    \centering
    \includegraphics[width=0.45\textwidth]{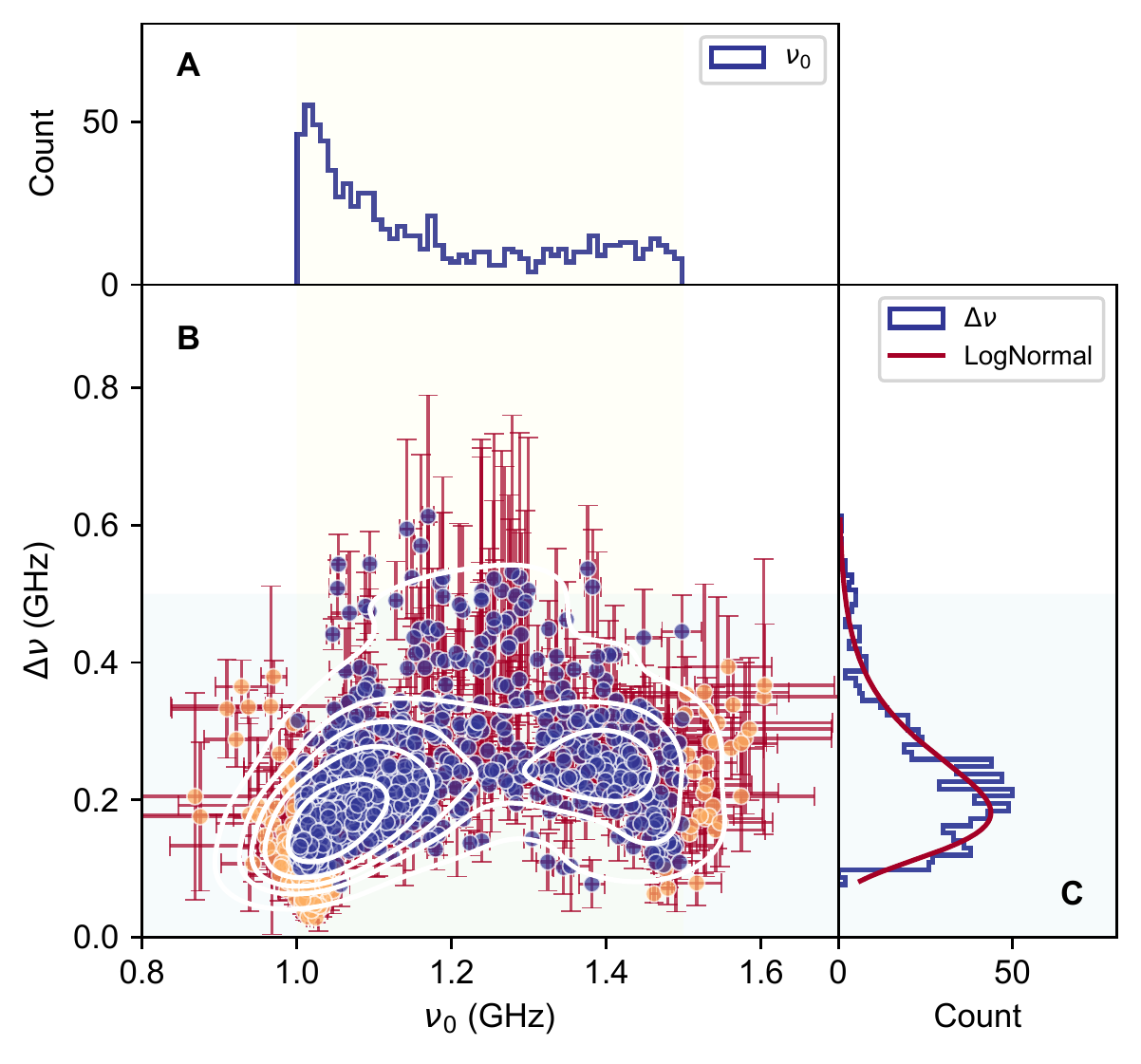} %
    \caption{Distribution of the center frequency ($\nu_0$) and spectral width ($\Delta\nu$) of the FRB~20220912A bursts fitted by Gaussian functions. The blue dots and lines in the figure represent bursts with peak frequency falling within the range of 1-1.5 GHz, and the fitting bandwidth exceeds 100 MHz when the peak frequency is below 1.05 GHz or above 1.45 GHz. The orange dots correspond to bursts that do not meet these conditions. \textbf{A}: The distribution of the center frequency. %
    \textbf{B}: The scatter plot of the center frequency and spectral width. The white contours show the 2D KDE illustrating the spatial distribution of the center frequency and spectral width. The light yellow and light blue backgrounds indicate the observed (1-1.5GHz) and effective (0.5GHz) bandwidths of the FAST telescope, respectively. \textbf{C}: The spectral width distribution fitted by a Log-Normal function with a peak value at 181 MHz.}
    \label{fig:freq}
\end{figure}

Figure~\ref{fig:freq} presents the distribution of the central frequency ($\nu_0$) and bandwidth ($\Delta\nu$) of the FRB~20220912A bursts. In order to mitigate the biases introduced by the limitations of observational bandwidth, we excluded bursts with $\nu_0$ falling outside the range of 1-1.5 GHzn, and bursts with $\nu_0$ within 50 MHz of the bandwidth edges (1 GHz and 1.5 GHz) with bandwidth $\Delta\nu < 100$ MHz, as these bursts are associated with high levels of fitting uncertainty. We subsequently analyzed the distribution of $\nu_0$ and $\Delta\nu$ for the remaining bursts. The central frequency of FRB~20220912A bursts mainly concentrates on the low-frequency range, and the distribution of high and low frequencies is extremely uneven. %
We use a single Log-Normal function to fit the bandwidth distribution. The mode of the Log-Normal function is located at 181 MHz, indicating that FRB~20220912A's bursts have very narrow-band spectra. Furthermore, we did not find significant correlation between central frequency and emission bandwidth according to the sample we collected at L-band. This is different from the power-law trend found by \cite{Kumar+23arXiv} from the bursts detected by the Parkes ultra-wideband lower receiver from the repeater FRB~20180301A, which might imply the actual observed bandwidth matters for this kind of analyses.

\begin{figure}[!htp]
    \centering
    \includegraphics[width=0.45\textwidth]{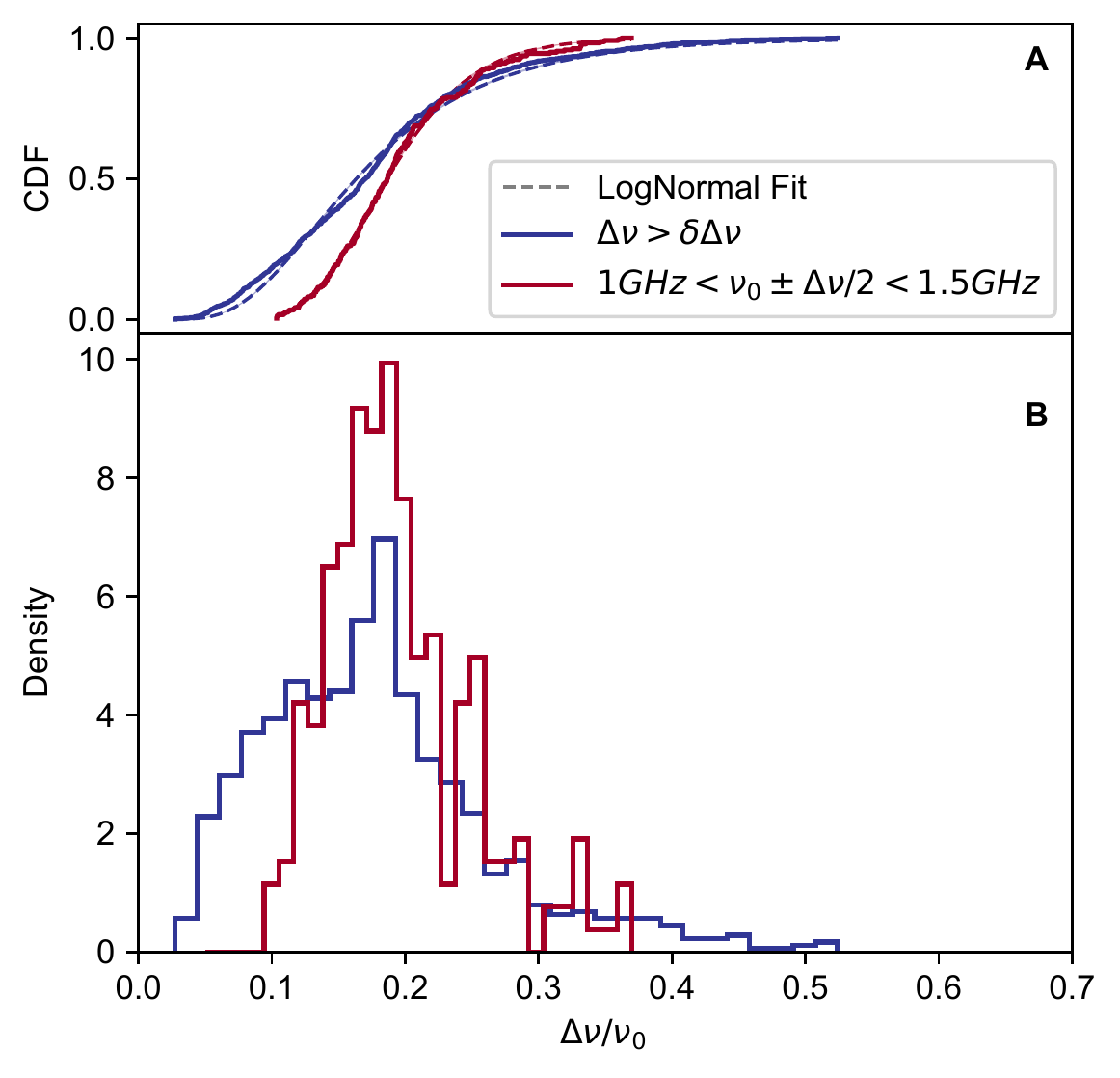}
    \caption{The $\Delta\nu/\nu_0$ distribution of FRB~20220912A. \textbf{A}: The cumulative distribution function of $\Delta\nu/\nu_0$. \textbf{B}: The differential probability distribution of $\Delta\nu/\nu_0$. For both panels, the blue lines and red lines are defined under two filtering criteria, i.e. the Gaussian fit standard deviation error $\delta\sigma$ is smaller than the standard deviation $\sigma$ itself (blue), and FWHM of the burst spectrum is completely within the range of 1-1.5 GHz (red).} %
    \label{fig:nuonu0}
\end{figure}

A more relevant parameter to describe the narrowness of an FRB spectrum is $\Delta\nu/\nu_0$. We investigate $\Delta\nu/\nu_0$ in more detail. Because many bursts have emission outside of the FAST band  (1-1.5 GHz), we limit our analysis to the bursts whose emission completely falls within the FAST band to avoid the bandwidth selection effect. We select bursts based on two stringent criteria, i.e. the Gaussian fit standard deviation error $\delta\sigma$ is smaller than the standard deviation $\sigma$ itself, and FWHM of the burst spectrum is completely within the range of 1-1.5 GHz. Figure~\ref{fig:nuonu0} displays the distribution $\delta \nu/\nu_0$ of these bursts under these two filtering criteria. %
We fit the hitogram distributions using lognormal functions and obtain $\mu=-1.812\pm0.001$ and $\sigma=0.475\pm0.002$ for blue histogram and $\mu=-1.679\pm0.001$ and $\sigma=0.265\pm0.002$ for red histogram, respectively. From the fitting results, we obtain their modes $e^{\mu-\sigma^2}$ located at 0.13 and 0.17, respectively. We also fit the cumulative $\Delta\nu/\nu_0$ distribution using a power-law function. The results are suboptimal with significant fitting errors.

\subsection{DM and Polarimetry} \label{sec:pol}

The DM of each burst was determined using the DM phase software package \footnote{\url{https://github.com/danielemichilli/DM_phase}}, which maximizes the coherent power in the pulse across the emission bandwidth. The median value of the DM is $220.70\,\rm pc\,cm^{-3}$ with a standard value is $1.83\,\rm pc\,cm^{-3}$, which is consistent with CHIME's measurement of $219.46\,\rm pc\,cm^{-3}$ \citep{2022ATel15679....1M}. Linear fitting indicates the absence of any trend in the time evolution of DM, with the slope is $d\,{\rm DM}/d\,{\rm t} < 
8\times10^{-3}\,\rm pc\,cm^{-3}\,day^{-1}$.

Due to the Faraday effect, the polarization plane of the FRB signal undergoes rotation during propagation. We employed the RM synthesis method to fit the RM of the burst using Eq.~\ref{eq:rm}, 

\begin{equation}\label{eq:rm}
\left(\begin{matrix}
I\\
Q'\\
U'\\
V
\end{matrix}\right) = \left[\begin{matrix}
1&0&0&0\\
0&\cos2\theta&\sin2\theta&0\\
0&-\sin2\theta&\cos2\theta&0\\
0&0&0&1\\
\end{matrix}\right]\left(\begin{matrix}
I\\
Q\\
U\\
V
\end{matrix}\right)
\end{equation}
where $\theta={\rm RM}\lambda^2$. We selected bursts with RM errors less than $10\,\rm rad\,m^{-2}$ (881 in total) for presentation.

The mean value of the RM is $-0.08\,\rm rad\,m^{-2}$, close to 0, indicating that the RM contribution from FRB~20220912A's host galaxy is comparable to that of the Milky Way, which is about $-16\,\rm rad\,m^{-2}$ \citep{2022A&A...657A..43H}. The low value of RM indicates that the FRB~20220912A may be in a very clean environment. The linear fit also suggests that there is no trend in the evolution of RM with time, with a slope of $0.017\pm0.018\ \rm day^{-1}$. Furthermore, several other active repeating FRBs, such as FRB~20121102A \citep{2021ApJ...908L..10H}, and 20201124A \citep{2022Natur.609..685X},  20190520B \citep{2022arXiv220308151D, 2022arXiv220211112A},  20180916B \citep{2022arXiv220509221M} and several other bursts (20181030A, 20181119A, 20190117A, 20190208A, 20190303A, 20190417A, \citealt{2023arXiv230208386M}) exhibit large RM values and show variations in RM on the timescale of months.
In contrast to these repeating FRBs with large RMs which suggest that all repeaters may have an associated synchrotron-emitting persistent radio sources (a supernova remnant, a magnetar wind nebula or a mini-AGN) with a dense and highly magnetized environment, the polarization data of FRB 20220912A suggests that a large and varying RM is not the necessary condition to make active repeaters. 

The linear polarization that is measured can be subject to overestimation when noise is present. As a result, we employ the frequency-averaged total linear polarization that has been de-biased \citep{2001ApJ...553..341E},

\begin{equation} \label{eq:L_de-bias}
    L_{\rm de\mbox{-}bias} = \left\{\begin{aligned}
      & \sigma_I \sqrt{\left(\frac{L_i}{\sigma_I}\right)^2 - 1} & \quad \frac{L_i}{\sigma_I} > 1.57 \\
      & 0                                                       & \quad \frac{L_i}{\sigma_I} \le 1.57
    \end{aligned}\right.
\end{equation}
where $\sigma_I$ represents the off-pulse standard deviation of Stokes I, while $L_i$ is the frequency-averaged linear polarization that has been measured for time sample $i$. The degree of linear and circular polarization are calculated with

\begin{equation} \label{eq:polde}
    L = \frac{\sum_i L_{{\rm de\mbox{-}bias}, i}}{\sum_i I_i} \quad {\rm and} \quad V = \frac{\sum_i V_i}{\sum_i I_i}
\end{equation}
where $V_i$ is defined similarly to $L_i$. The uncertainties on the linear polarization fraction and circular polarization fraction are calculated as:

\begin{equation} \label{eq:poler}
    \sigma = \frac{\sigma_{I}}{I}\sqrt{N+N\left(\rho^2/I^2\right)}
\end{equation}
where $N$ is the number of time samples of the burst, and $\rho$ is $\sum_i L_{{\rm de\mbox{-}bias}, i}$ or $\sum_i V_i$, depending on whether we are calculating the linear or circular polarization fraction.

\begin{figure}[!htp]
    \centering
    \includegraphics[width=0.45\textwidth]{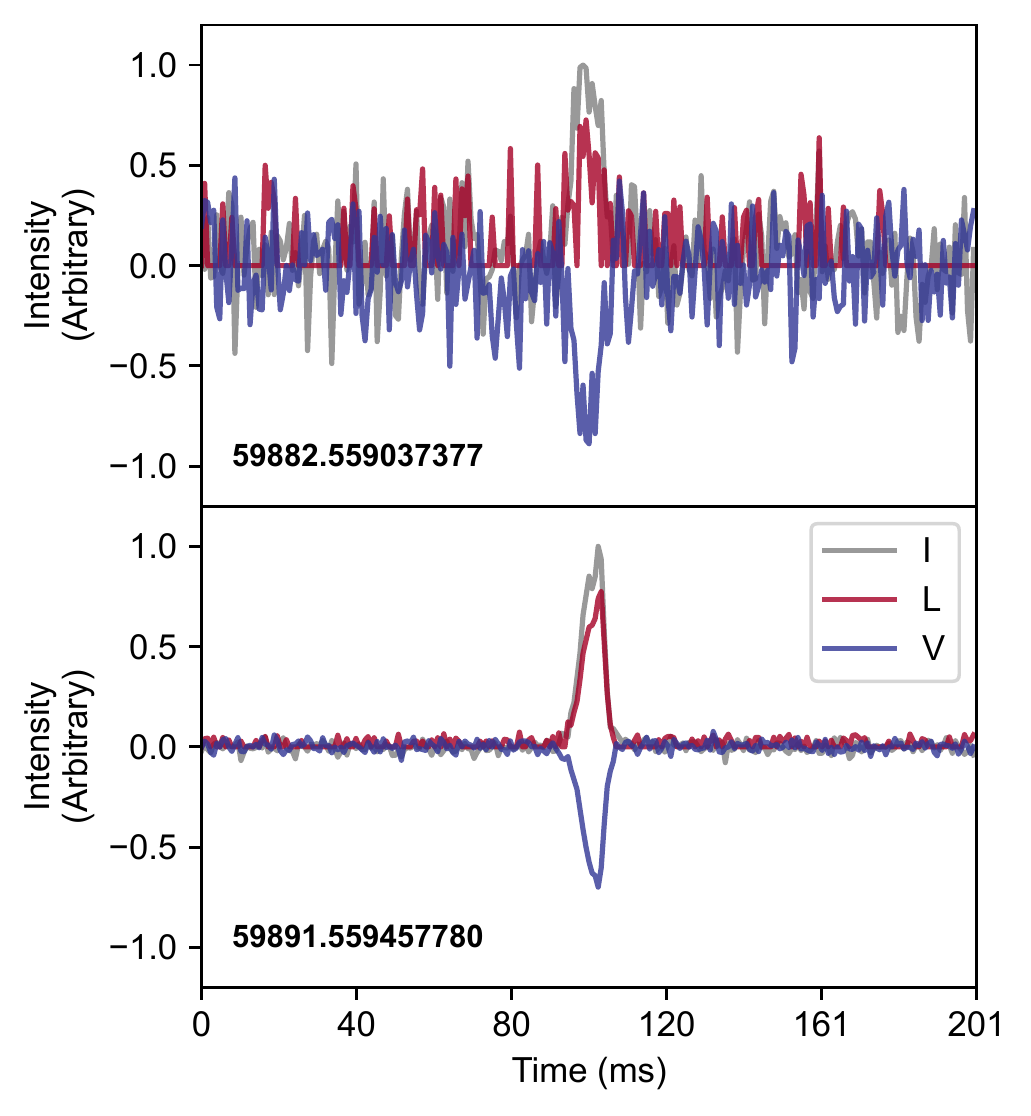}\\
    \caption{Two bursts with the highest circular polarization degrees. The black, red, and blue lines respectively represent the burst's total intensity, linear polarization, and circular polarization profiles.}
    \label{fig:cirmax}
\end{figure}

Most of the bursts from FRB~20220912A exhibit almost 100\% linear polarization, with a noticeable fraction of bursts exhibiting significant circular polarization. Figure~\ref{fig:cirmax} displays the two bursts with the highest circular polarization degrees, which are $-78.0\pm10.8\,\%$ and $-69.4\pm1.4\,\%$, respectively.

\begin{figure*}[!htp]
    \centering
    \includegraphics[width=0.9\textwidth]{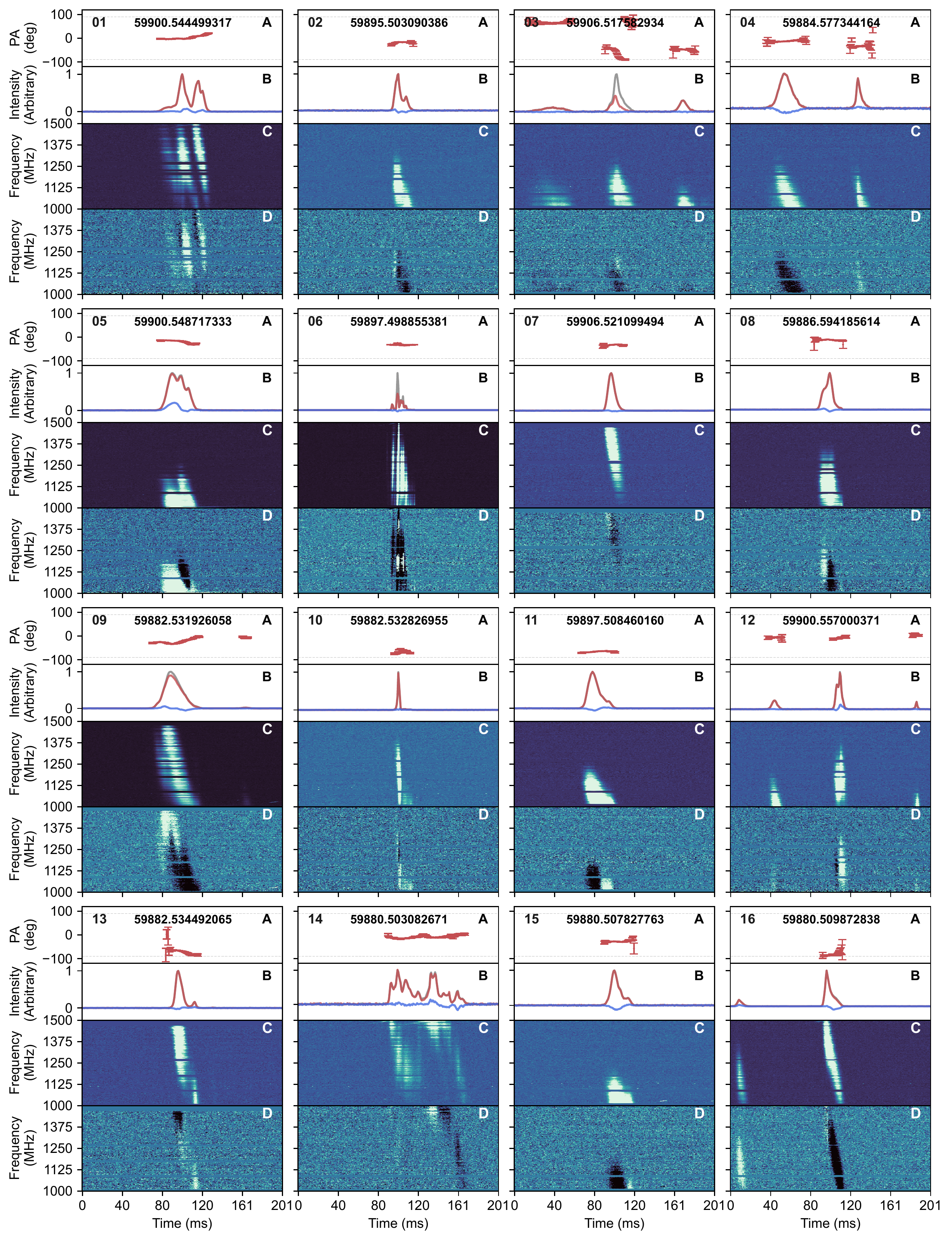}\\
    \caption{16 bright bursts from FRB~20220912A with circular polarization reversal over time or/and frequency. \textbf{A}: polarization position angle (PA). \textbf{B}: polarization profiles of the bursts, with the black, red, and blue lines representing the total intensity, linear polarization, and circular polarization, respectively. \textbf{C}: Stokes I. \textbf{D}: Stokes V.}
    \label{fig:cirvaryf}
\end{figure*}

Additionally, the circular polarization dynamic spectra of FRB~20220912A show various morphologies. We display 16 bursts in Figure~\ref{fig:cirvaryf}. It can be seen that some bursts that cannot be distinguished in Stokes I appear as multiple bursts in Stokes V, such as bursts 1, 2, 7, 11, and 16.
The bursts that show sign changes in circular polarization also have sub-pulse structures, so that the sign change may be caused by sub-pulses with different circular polarization modes at different times.
The mechanism responsible for circular polarization is still under extensive discussion and no definitive conclusion has been reached \citep{2023arXiv230209697Q}.
Curvature radiation by emitting bunches can be circularly polarized if the line of sight  (LOS) is not confined into beaming angle \citep{2022ApJ...927..105W}. A sign change of circular polarization can be seen if the opening angle of bunch is not much larger than $1/\gamma$, where $\gamma$ is the Lorentz factor of the bunch \citep{2022MNRAS.517.5080W}. The curvature radiation mechanism predicts an average circular polarization fraction smaller than $55\%$ when a sign change of circular polarization occurs. 
Another intrinsic radiation mechanism is the coherent inverse Compton scattering through charged bunches \citep{2022ApJ...925...53Z, 2023MNRAS.518...66Q}. The scattered waves can be circularly polarized by adding up linearly polarized waves with different phases and polarization angles \citep{2023arXiv230209697Q}. When the LOS sweeps across the bunch central axis, the sign of circular polarization can change and the maximum circular polarization fraction can be larger than that of curvature radiation. The observation of FRB 20220912A could be understood within either scenario. 

\begin{figure}[!htp]
    \centering
    \includegraphics[width=0.45\textwidth]{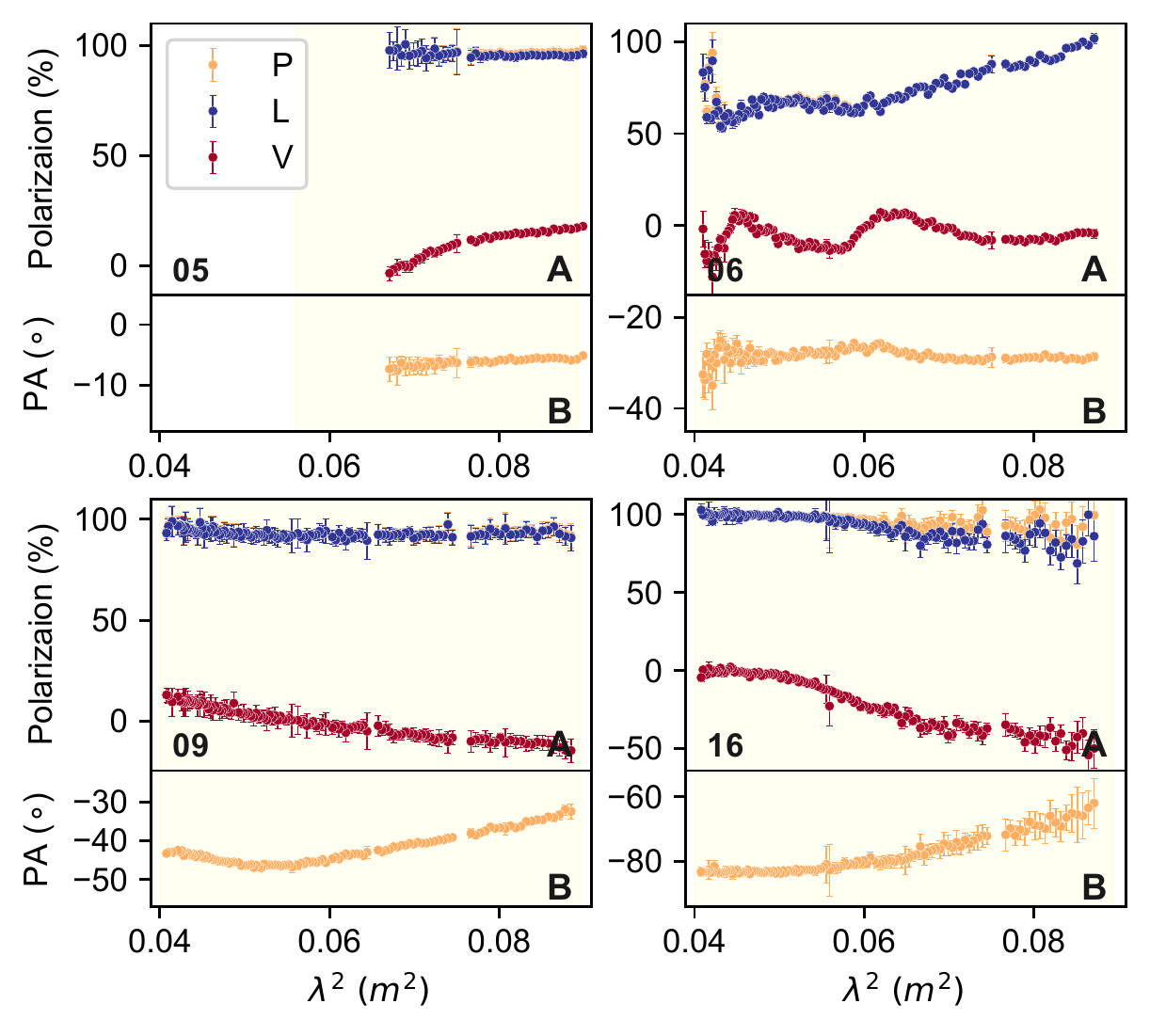}\\
    \caption{4 bright bursts from FRB~20220912A with circular polarization oscillating over frequency. \textbf{A}: polarization degree. \textbf{B}: PA.}
    \label{fig:cirfreq}
\end{figure}

Circular polarization in some bursts also appears to vary with frequency
(Figure~\ref{fig:cirfreq}). An oscillation of polarization parameters as a function of wavelength may be an indication of Faraday conversion \citep{2022Natur.609..685X,2023arXiv230209697Q}.
However, the polarization degree of FRB~20220912A oscillates without an obvious oscillation frequency like FRB~20201124A \citep{2022Natur.609..685X}. Furthermore, the conditions for significant Faraday conversion are usually quite stringent \citep{2023arXiv230209697Q}, requiring special magnetic field reversal environments, 
e.g. in a binary star system \citep{2022NatCo..13.4382W} or when the source is surrounded by a supernova remnant \citep{2023MNRAS.520.2039Y}. Such a required environment is not consistent with the very clean environment inferred from the small and non-variable RM of FRB~20220912A.

\section{Discussion} \label{sec:disc}

\subsection{Total Energy Budget}

By utilizing the isotropic energies derived from each burst, it is possible to impose limitations on the total energy allocation of the intrinsic FRB source, which can be employed to restrict the various models of FRB sources. When deducing the total source energy based on the energy of each burst, various factors must be considered, including radio radiation efficiency $\eta_r$, beaming factor $f_b$, and observation duty cycle $\zeta$ \citep{zhang22}. 

We use Eq.~\ref{eq:energy} for energy calculation with the isotropic emission assumption. However, coherent radiation from FRB bursts generally has a small solid angle $\delta\Omega$. Therefore, the true burst energy should be $\delta\Omega/4\pi$ of the isotropic burst energy. Additionally, bursts may exist in directions that are not observable. Assuming the global emission beam to be $\Delta\Omega$, we can get the global beaming factor $f_b=\Delta\Omega/4\pi$. 

Due to the limitation of the observation duty cycle, we cannot observe all bursts emitted by the FRB source during the observation period. For example, our observations of FRB~20220912A were conducted over 17 days with a total observation time of only 8.67 hours, which does not imply that there was no activity at other hours. Therefore, to estimate the total source energy during the observation period, we can use duty cycle scaling to compensate for the unobserved periods of FRB radiation energy. If the total radio energy of the observed bursts is $E_b$, the total source energy should be $E_b\times F_b\times\eta_r^{-1}\times\zeta^{-1}$.

Table~\ref{tab:energycal} presents our estimates of the total source energy for FRB~20121102A \citep{2021Natur.598..267L}, FRB~20190520B \citep{2022Natur.606..873N}, FRB~20201124A \citep{2022Natur.609..685X, 2022RAA....22l4002Z}, and FRB~20220912A (this paper) using FAST observations. In our calculations, we assumed typical values for $\eta_r$ and $F_b$ of $10^{-4}$ and 0.1, respectively. It should be noted that the burst energies reported in the literature for FRB~20121102A and FRB~20190520B were calculated using a center frequency of 1.25~GHz, rather than the bandwidth $\Delta\nu$. Therefore, we have divided the total energies of these two FRBs by 2.5, to allow for comparison with the other FRBs (see footnote 1 %
for more discussion). FRB~20220912A emitted an energy of $3.49\times10^{45}$ during the 17-day period, which exceeds 2\% of the total dipolar magnetic energy ($E_B\sim1.7\times10^{47}$ erg) of a magnetar. This implies that the dipolar magnetic energy of the magnetar would be completely depleted in only $\sim$ 850 days if the radio efficiency is indeed as low as $10^{-4}$. Certain magnetar models (e.g. low-efficiency models invoking relativistic shocks) would suffer from an energy budget problem. 

\begin{deluxetable}{ccccc}
\tablecaption{Energy budget of four repeating FRBs.\label{tab:energycal}}
\setlength\tabcolsep{2pt}
\tabletypesize{\scriptsize}
\tablehead{\\[-5pt]
 FRB Name & Duty Cycle$^a$ & Radio E$^b$ & Averaged E$^c$  & Source E$^d$  \\
          & $\zeta$        & (erg)       & (erg hr$^{-1}$) & ($\eta_{r,-4}^{-1}F_{b,-1}$ erg)$^e$              
}
\startdata
  20121102A     & 0.053  & $1.36\times10^{41}$ & $2.29\times10^{39}$ & $2.59\times10^{45}$ \\
  20190520B     & 0.070  & $4.39\times10^{39}$ & $2.37\times10^{38}$ & $6.26\times10^{43}$ \\
  20201124A$^f$ & 0.063  & $1.65\times10^{41}$ & $2.01\times10^{39}$ & $2.60\times10^{45}$ \\
  20201124A$^g$ & 0.042  & $6.42\times10^{40}$ & $1.60\times10^{40}$ & $1.54\times10^{45}$ \\
  20220912A     & 0.021  & $7.42\times10^{40}$ & $8.55\times10^{39}$ & $3.49\times10^{45}$ \\
\enddata
\tablecomments{$^a$ The observation duty cycle, e.g. for FRB~20220912A in this paper, the duty cycle is 8.67 hours out of 17 days. $^b$ Sum of the observed isotropic radio energies of all bursts. $^c$ The total radio energy divided by observation time, e.g. for FRB~20220912A in this paper, the averaged energy is $7.42\times10^{40}$ erg / 8.67 hours. $^d$ The total source energy. $^e$ The source energy calculation uses $\eta_r=10^{-4}$ and $F_b=0.1$. $^f$ FAST observation of FRB~20201124A in 2021.04 by \cite{2022Natur.609..685X}. $^g$ FAST observation of FRB~20201124A in 2021.09 by \cite{2022RAA....22l4002Z}}.
\end{deluxetable}

\subsection{Circular Polarization and Environment}

FRB~20201124A was the first repeating FRB discovered with circular polarization \citep{2021MNRAS.508.5354H}. Prior to this, only some non-repeating bursts were found to have significant circular polarization \citep{2020ApJ...891L..38C, 2020MNRAS.497.3335D, 2022Sci...375.1266F}. Recently, FRB~20121102A and FRB~20190520B were detected by FAST with very few bursts exhibiting circular polarization \citep{2022SciBu..67.2398F}. Here, we present that FRB~20220912A has a large number of bursts exhibiting circular polarization, suggesting that this may be a common feature of repeating FRBs. 

All the above-mentioned four repeating FRBs have  a large number of detected bursts by FAST. We counted the proportion of bursts exhibiting significant circular polarization (circular degree $\ge 10\%$) for each of these four repeating FRBs. FRB~20121102A exhibited 12 out of 1652 detected bursts with circular polarization \citep{2022SciBu..67.2398F, 2021Natur.598..267L}. Similarly, FRB~20190520B displayed circular polarization in 3 out of 75 detected bursts \citep{2022SciBu..67.2398F, 2022Natur.606..873N}. FRB~20201124 showed circular polarization in 302 out of 1863 detected bursts \citep{2022Natur.609..685X}, while FRB~20220912A exhibited circular polarization in 303 out of 1076 detected bursts (this paper).

\begin{figure}[!htp]
    \centering
    \includegraphics[width=0.45\textwidth]{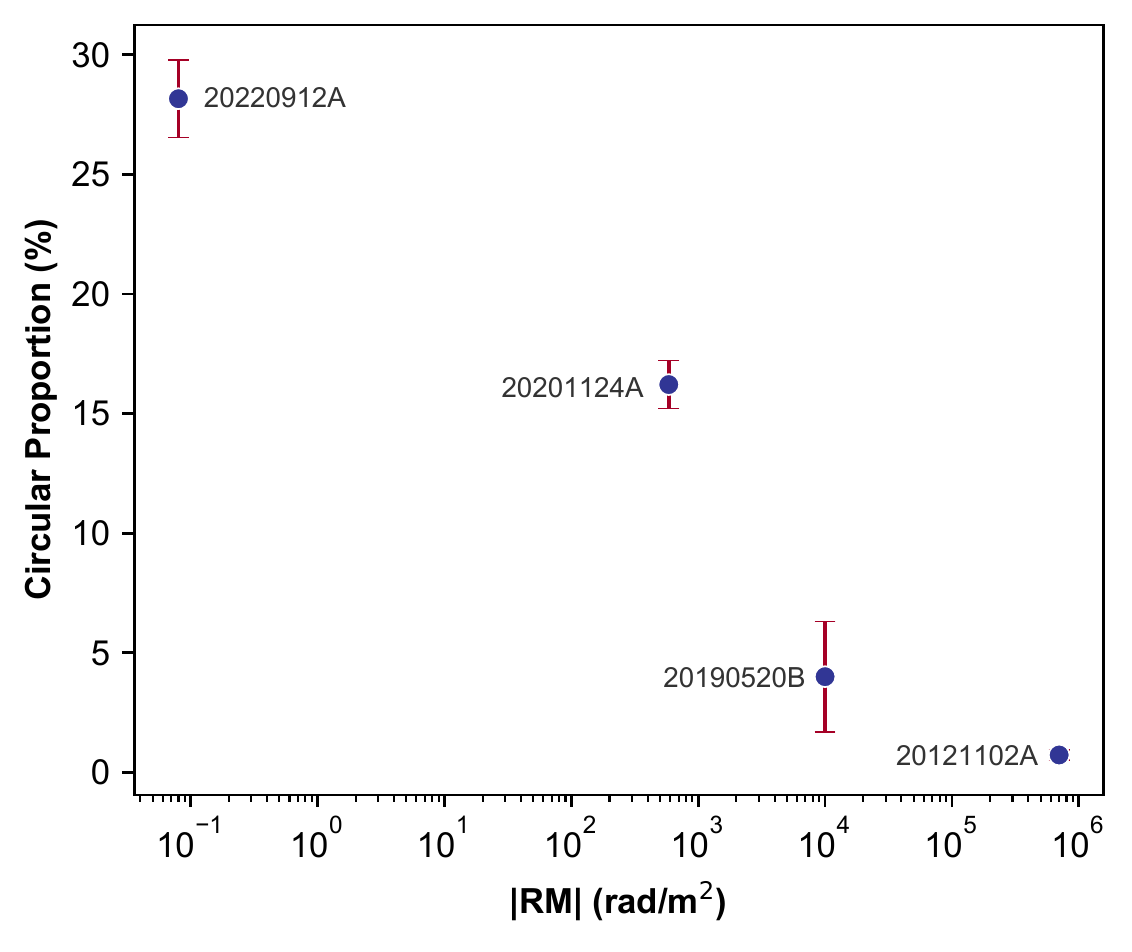}
    \caption{The proportion of bursts with circular polarization from repeating FRBs varies with RM. The red error bars are given by the Poisson counting error of the statistical counts. The error bar on FRB~20121102A is smaller than the symbol size.}
    \label{fig:cp}
\end{figure}

Figure~\ref{fig:cp} shows a tentative relationship between the fraction of bursts with circular polarization degree $>10\%$ and the absolute value of each FRB's RM. 
It appears that there is a negative correlation. The larger the RM value, the lower of the fraction of bursts with circular polarization. If such a correlation is physical, it may be related to the intrinsic radiation mechanism and complicated environments of repeating FRB sources. 
First, the larger fraction of circular polarization in the clean-environment FRB~20121102A suggests that the origin of circular polarization may be more related to intrinsic radiation mechanisms than  environment effects. According to \cite{2023arXiv230209697Q}, significant circular polarization can be made from magnetospheric radiation mechanisms such as coherent curvature radiation or inverse Compton scattering by bunches. Relativistic shock models invoking synchrotron maser radiation mechanism, on the other hand, mostly emit bursts with nearly 100\% linear polarization. The detection of circular polarization from all four sources favors the magnetospheric origin of the bursts. Second For both curvature and inverse Compton scattering processes, given a random viewing angle the fraction of bursts that have high circular polarization degree is high if the bunch shape is point like. In order to reduce the fraction of circularly polarized bursts, the cross section of the bunches should be large so that most emitting leptons are viewed on beam \citep[e.g.][]{2022MNRAS.517.5080W}. The correlation seen in Figure~\ref{fig:cp} would then require that the FRB engine in a more magnetized environment (e.g. a younger magnetar) should be able to generate bunches with larger cross sections. Such a scenario may predict the the bursts from a more magnetized environment are systematically brighter, which is not observed. An alternative, possibly more plausible possibility is that the circular polarization fraction is modified by environments. \cite{2023arXiv230209697Q} showed that polarization-mode-selected synchrotron absorption tends to convert circular polarization to linear polarization. The observed trend in Figure~\ref{fig:cp} then suggests that there are more significant synchrotron absorption in more magnetized environments. More repeater data are needed to confirm whether the circular polarization fraction and RM correlation is physical.

\section{Conclusions} \label{sec:conclusion}

We report the observation of FRB~20220912A by FAST in 2022. 

\begin{itemize}
\setlength{\itemsep}{3pt}

    \item[1] The observation was conducted 17 times for a total of 8.67 hours.
    
    \item[2] A total of 1076 bursts were detected, with the highest event rate of 390 hr$^{-1}$. No periodicity in the range of 1 ms--1000 s or 2--27 d was detected.
    
    \item[3] The energy distribution of FRB~20220912A cannot be described by a single function. The differential energy distribution was described using two Log-Normal functions, with characteristic energies $5.29\times10^{36}$ erg and $4.13\times10^{37}$ erg. The cumulative energy distribution was described using a broken power-law function, with power-law indices $-0.38\pm0.02$ and $-2.07\pm0.07$.
    
    \item[4] We report, for the first time, the synthetic spectrum of many bursts of FRB~20220912A. It can be fitted by a power law function with an L-band spectral index of -2.6.
    
    \item[5] The RM of FRB~20220912A is close to 0 and did not show any evolution during the two-month observation period, indicating that the contribution of the Milky Way and the host galaxy to RM is comparable. The contribution of the Milky Way is estimated to be $-16\,\rm rad\,m^{-2}$, suggesting that FRB~20220912A is located in a relatively clean local environment. This burst suggest that a high and variable RM is not the necessary condition of making an active FRB repeater. 
    
    \item[6] Most bursts of FRB~20220912A in the L band exhibit nearly 100\% linear polarization and a large fraction of the bursts exhibit circular polarization, with a maximum of 70\%. Some of the bursts with circular polarization changing sign over time or frequency. The high circular polarization degree of the bursts is likely related to the intrinsic radiation mechanisms. Likely models include coherent curvature radiation and inverse Compton scattering by bunches within the magnetosphere of the FRB source. 
    
    \item[7] We found a trenative negative relation between $\rm |RM|$ and the circular polarization fraction, i.e. a larger $\rm |RM|$ corresponds to a smaller fraction of circularly polarized bursts. If such a correlation is real, it may imply more mode-selected synchrotron absorption in more magnetized environments.

\end{itemize}

Multi-wavelength observations of FRBs are essential to understanding their origins and environments. FRB~20220912A is located in a host galaxy at with $z\sim0.077$, making it the second closest repeating FRBs after FRB~20200120E \citep{2022Natur.602..585K}. Its proximity means that multi-wavelength observations will be more efficient. Furthermore, FRB~20220912A is a very active repeating FRB, which makes it an ideal target for multi-wavelength observations. We encourage further multi-wavelength observations of FRB~20220912A to unravel the mystery of FRBs.

\section*{Data Availability}
The full table for bursts' properties is in Table~\ref{tab:burst}. The data underlying this paper will be shared on a reasonable request to the corresponding author.

\acknowledgments
This work made use of the data from FAST, a Chinese national mega-science facility, operated by National Astronomical Observatories, Chinese Academy of Sciences. This work was supported by the National Natural Science Foundation of China (NSFC, Grant No. 11988101, 12203045, 11725313, 12003028 and 12041303).
This work was supported by the National SKA Program of China No. 2020SKA0120200 and 2022SKA0130100, the CAS-MPG LEGACY project, and YSBR-063, CAS Project for Young Scientists in Basic Research. 

\appendix
\section{Burst Property}

\movetabledown=57mm
\begin{rotatetable}
\begin{deluxetable*}{cccccccccccc}
\tablecaption{The properties of the FRB~20220912A bursts\label{tab:burst}}
\tablewidth{700pt}
\tabletypesize{\small}
\tablehead{
 Burst ID & MJD$^a$ & DM             & PeakFlux$^b$  & Width$^c$      & PeakFrequency$^d$ & BandWidth$^e$ & Fluence$^b$ & Energy     & RM             & Linear & Circular \\
          &         & (pc cm$^{-3}$) & (mJy)         & (ms)           & (MHz)             & (MHz)         & (Jy ms)     & (erg)      & (rad m$^{-2}$) & (\%)   & (\%)
}
\startdata
B01 & 59880.497624400   & 220.24$\pm$1.76   & 117.4$\pm$1.4   & 4.70   & 1479.5$\pm$5.4    & 122.4$\pm$9.3     & 0.552$\pm$0.007   & 3.979(49)e+37    & -4.3$^{+2.1}_{-1.9}$   & 95.0$\pm$1.4   & 5.3$\pm$1.0 \\
B02 & 59880.497624673   & 217.82$\pm$2.9    & 115.5$\pm$1.4   & 3.72   & 1366.2$\pm$4.3    & 231.4$\pm$15.0    & 0.429$\pm$0.005   & 3.097(38)e+37    & -0.3$^{+0.8}_{-0.8}$   & 98.5$\pm$1.8   & -16.9$\pm$1.3 \\
B03 & 59880.497684224   & 221.87$\pm$0.06   & 242.8$\pm$3.0   & 4.09   & 1300.0$\pm$13.7   & 735.7$\pm$896.2   & 0.993$\pm$0.012   & 7.161(88)e+37    & 1.1$^{+0.2}_{-0.2}$    & 97.7$\pm$1.0   & 3.3$\pm$0.7 \\
B04 & 59880.497945389   & 220.88$\pm$3.78   & 174.6$\pm$2.1   & 3.54   & 1381.0$\pm$6.8    & 321.4$\pm$28.0    & 0.619$\pm$0.008   & 4.463(55)e+37    & 0.5$^{+0.5}_{-0.5}$    & 98.6$\pm$1.4   & -6.7$\pm$1.0 \\
B05 & 59880.497991706   & 220.91$\pm$0.24   & 32.1$\pm$0.4    & 4.06   & 1095.7$\pm$11.2   & 251.7$\pm$34.6    & 0.130$\pm$0.002   & 9.398(115)e+36   & -3.1$^{+1.5}_{-1.8}$   & 94.5$\pm$6.7   & -16.1$\pm$4.9 \\
B06 & 59880.498128495   & 219.64$\pm$5.08   & 95.2$\pm$1.2    & 4.89   & 1564.0$\pm$56.7   & 338.4$\pm$65.8    & 0.465$\pm$0.006   & 3.355(41)e+37    & 2.1$^{+1.2}_{-1.2}$    & 90.6$\pm$1.6   & 7.2$\pm$1.2 \\
B07 & 59880.498199601   & 223.30$\pm$0.17   & 138.2$\pm$1.7   & 8.97   & 1449.5$\pm$26.3   & 332.1$\pm$63.9    & 1.240$\pm$0.015   & 8.942(109)e+37   & 0.1$^{+0.4}_{-0.3}$    & 97.2$\pm$0.9   & 14.8$\pm$0.6 \\
B08 & 59880.498605364   & 222.98$\pm$1.36   & 96.5$\pm$1.2    & 8.65   & 1399.0$\pm$15.2   & 374.2$\pm$61.9    & 0.834$\pm$0.01    & 6.019(73)e+37    & 3.1$^{+0.6}_{-0.6}$    & 97.8$\pm$1.5   & -2.7$\pm$1.0 \\
B09 & 59880.498635029   & 220.22$\pm$0.24   & 34.5$\pm$0.4    & 0.64   & 1004.4$\pm$22.4   & 110.3$\pm$34.3    & 0.022$\pm$0.0     & 1.592(19)e+36    & -                      & -              & - \\
B10 & 59880.498915641   & 221.35$\pm$0.21   & 77.7$\pm$0.9    & 8.74   & 1083.0$\pm$5.5    & 256.5$\pm$16.1    & 0.679$\pm$0.008   & 4.898(59)e+37    & 1.9$^{+0.6}_{-0.6}$    & 98.7$\pm$2.2   & -3.8$\pm$1.6 \\
\enddata
\tablecomments{$^a$ Barycentrical arrival time at 1.5 GHz\\
$^b$ Calculated within 500 MHz bandwidth\\
$^c$ Equivalent width\\
$^d$ Obtained with Gaussian fitting\\
$^e$ FWHM of Gaussian fitting\\
The full table will be available in ScienceDB \url{https://doi.org/10.57760/sciencedb.08058}.}
\end{deluxetable*}
\end{rotatetable}

\bibliographystyle{aasjournal}
\bibliography{ref}{}

\begin{thebibliography}{}
\expandafter\ifx\csname natexlab\endcsname\relax\def\natexlab#1{#1}\fi
\providecommand{\url}[1]{\href{#1}{#1}}
\providecommand{\dodoi}[1]{doi:~\href{http://doi.org/#1}{\nolinkurl{#1}}}
\providecommand{\doeprint}[1]{\href{http://ascl.net/#1}{\nolinkurl{http://ascl.net/#1}}}
\providecommand{\doarXiv}[1]{\href{https://arxiv.org/abs/#1}{\nolinkurl{https://arxiv.org/abs/#1}}}

\bibitem[{{Aggarwal} {et~al.}(2021){Aggarwal}, {Agarwal}, {Lewis},
  {Anna-Thomas}, {Tremblay}, {Burke-Spolaor}, {McLaughlin}, \&
  {Lorimer}}]{2021ApJ...922..115A}
{Aggarwal}, K., {Agarwal}, D., {Lewis}, E.~F., {et~al.} 2021, \apj, 922, 115,
  \dodoi{10.3847/1538-4357/ac2577}

\bibitem[{{Anna-Thomas} {et~al.}(2022){Anna-Thomas}, {Connor}, {Burke-Spolaor},
  {Beniamini}, {Aggarwal}, {Law}, {Lynch}, {Li}, {Feng}, {Ocker}, {Cruces},
  {Chatterjee}, {Yu}, {Niu}, \& {Xue}}]{2022arXiv220211112A}
{Anna-Thomas}, R., {Connor}, L., {Burke-Spolaor}, S., {et~al.} 2022, arXiv
  e-prints, arXiv:2202.11112, \dodoi{10.48550/arXiv.2202.11112}

\bibitem[{{Bhusare} {et~al.}(2022){Bhusare}, {Kumar}, {Maan}, {Marthi},
  {Tendulkar}, {Lal}, {Lin}, {Main}, \& {Pal}}]{2022ATel15806....1B}
{Bhusare}, Y., {Kumar}, A., {Maan}, Y., {et~al.} 2022, The Astronomer's
  Telegram, 15806, 1

\bibitem[{{Chatterjee} {et~al.}(2017){Chatterjee}, {Law}, {Wharton},
  {Burke-Spolaor}, {Hessels}, {Bower}, {Cordes}, {Tendulkar}, {Bassa},
  {Demorest}, {Butler}, {Seymour}, {Scholz}, {Abruzzo}, {Bogdanov}, {Kaspi},
  {Keimpema}, {Lazio}, {Marcote}, {McLaughlin}, {Paragi}, {Ransom}, {Rupen},
  {Spitler}, \& {van Langevelde}}]{2017Natur.541...58C}
{Chatterjee}, S., {Law}, C.~J., {Wharton}, R.~S., {et~al.} 2017, \nat, 541, 58,
  \dodoi{10.1038/nature20797}

\bibitem[{{Chime/Frb Collaboration} {et~al.}(2020){Chime/Frb Collaboration},
  {Amiri}, {Andersen}, {Bandura}, {Bhardwaj}, {Boyle}, {Brar}, {Chawla},
  {Chen}, {Cliche}, {Cubranic}, {Deng}, {Denman}, {Dobbs}, {Dong}, {Fandino},
  {Fonseca}, {Gaensler}, {Giri}, {Good}, {Halpern}, {Hessels}, {Hill},
  {H{\"o}fer}, {Josephy}, {Kania}, {Karuppusamy}, {Kaspi}, {Keimpema},
  {Kirsten}, {Landecker}, {Lang}, {Leung}, {Li}, {Lin}, {Marcote}, {Masui},
  {McKinven}, {Mena-Parra}, {Merryfield}, {Michilli}, {Milutinovic},
  {Mirhosseini}, {Naidu}, {Newburgh}, {Ng}, {Nimmo}, {Paragi}, {Patel}, {Pen},
  {Pinsonneault-Marotte}, {Pleunis}, {Rafiei-Ravandi}, {Rahman}, {Ransom},
  {Renard}, {Sanghavi}, {Scholz}, {Shaw}, {Shin}, {Siegel}, {Singh}, {Smegal},
  {Smith}, {Stairs}, {Tendulkar}, {Tretyakov}, {Vanderlinde}, {Wang}, {Wang},
  {Wulf}, {Yadav}, \& {Zwaniga}}]{2020Natur.582..351C}
{Chime/Frb Collaboration}, {Amiri}, M., {Andersen}, B.~C., {et~al.} 2020, \nat,
  582, 351, \dodoi{10.1038/s41586-020-2398-2}

\bibitem[{{Chime/Frb Collaboration} {et~al.}(2022){Chime/Frb Collaboration},
  {Bandura}, {Bhardwaj}, {Boyle}, {Brar}, {Breitman}, {Cassanelli},
  {Chatterjee}, {Chawla}, {Cliche}, {Cubranic}, {Curtin}, {Deng}, {Dobbs},
  {Dong}, {Fonseca}, {Gaensler}, {Giri}, {Good}, {Hill}, {Josephy},
  {Kaczmarek}, {Kader}, {Kania}, {Kaspi}, {Leung}, {Li}, {Lin}, {Masui},
  {McKinven}, {Mena-Parra}, {Merryfield}, {Meyers}, {Michilli}, {Naidu},
  {Newburgh}, {Ng}, {Ordog}, {Patel}, {Pearlman}, {Pen}, {Petroff}, {Pleunis},
  {Rafiei-Ravandi}, {Rahman}, {Ransom}, {Renard}, {Sanghavi}, {Scholz}, {Shaw},
  {Shin}, {Siegel}, {Singh}, {Smith}, {Stairs}, {Tan}, {Tendulkar},
  {Vanderlinde}, {Wiebe}, {Wulf}, \& {Zwaniga}}]{2022Natur.607..256C}
{Chime/Frb Collaboration}, Andersen, B.~C., {Bandura}, K., {Bhardwaj}, M.,
  {et~al.} 2022, \nat, 607, 256, \dodoi{10.1038/s41586-022-04841-8}

\bibitem[{{Cho} {et~al.}(2020){Cho}, {Macquart}, {Shannon}, {Deller},
  {Morrison}, {Ekers}, {Bannister}, {Farah}, {Qiu}, {Sammons}, {Bailes},
  {Bhandari}, {Day}, {James}, {Phillips}, {Prochaska}, \&
  {Tuthill}}]{2020ApJ...891L..38C}
{Cho}, H., {Macquart}, J.-P., {Shannon}, R.~M., {et~al.} 2020, \apjl, 891, L38,
  \dodoi{10.3847/2041-8213/ab7824}

\bibitem[{{Cordes} \& {Lazio}(2002)}]{2002astro.ph..7156C}
{Cordes}, J.~M., \& {Lazio}, T.~J.~W. 2002, arXiv e-prints, astro.
\newblock \doarXiv{astro-ph/0207156}

\bibitem[{{Cruces} {et~al.}(2021){Cruces}, {Spitler}, {Scholz}, {Lynch},
  {Seymour}, {Hessels}, {Gouiff{\'e}s}, {Hilmarsson}, {Kramer}, \&
  {Munjal}}]{2021MNRAS.500..448C}
{Cruces}, M., {Spitler}, L.~G., {Scholz}, P., {et~al.} 2021, \mnras, 500, 448,
  \dodoi{10.1093/mnras/staa3223}

\bibitem[{{Dai} {et~al.}(2022){Dai}, {Feng}, {Yang}, {Zhang}, {Li}, {Niu},
  {Wang}, {Xue}, {Zhang}, {Burke-Spolaor}, {Law}, {Lynch}, {Connor},
  {Anna-Thomas}, {Zhang}, {Duan}, {Yao}, {Tsai}, {Zhu}, {Cruces}, {Hobbs},
  {Miao}, {Niu}, {Filipovic}, \& {Zhu}}]{2022arXiv220308151D}
{Dai}, S., {Feng}, Y., {Yang}, Y.~P., {et~al.} 2022, arXiv e-prints,
  arXiv:2203.08151, \dodoi{10.48550/arXiv.2203.08151}

\bibitem[{{Day} {et~al.}(2020){Day}, {Deller}, {Shannon}, {Qiu(邱昊)},
  {Bannister}, {Bhandari}, {Ekers}, {Flynn}, {James}, {Macquart}, {Mahony},
  {Phillips}, \& {Xavier Prochaska}}]{2020MNRAS.497.3335D}
{Day}, C.~K., {Deller}, A.~T., {Shannon}, R.~M., {et~al.} 2020, \mnras, 497,
  3335, \dodoi{10.1093/mnras/staa2138}

\bibitem[{Dunning {et~al.}(2017)Dunning, Bowen, Castillo, Chung, Doherty,
  George, Hayman, Jeganathan, Kanoniuk, Mackay, {et~al.}}]{FAST19Beam}
Dunning, A., Bowen, M., Castillo, S., {et~al.} 2017, in 2017 XXXIInd General
  Assembly and Scientific Symposium of the International Union of Radio Science
  (URSI GASS), IEEE, 1--4

\bibitem[{{Everett} \& {Weisberg}(2001)}]{2001ApJ...553..341E}
{Everett}, J.~E., \& {Weisberg}, J.~M. 2001, \apj, 553, 341,
  \dodoi{10.1086/320652}

\bibitem[{{Fedorova} \& {Rodin}(2022)}]{2022ATel15713....1F}
{Fedorova}, V.~A., \& {Rodin}, A.~E. 2022, The Astronomer's Telegram, 15713, 1

\bibitem[{{Feng} {et~al.}(2022{\natexlab{a}}){Feng}, {Zhang}, {Li}, {Yang},
  {Wang}, {Niu}, {Dai}, \& {Yao}}]{2022SciBu..67.2398F}
{Feng}, Y., {Zhang}, Y.-K., {Li}, D., {et~al.} 2022{\natexlab{a}}, Science
  Bulletin, 67, 2398, \dodoi{10.1016/j.scib.2022.11.014}

\bibitem[{{Feng} {et~al.}(2022{\natexlab{b}}){Feng}, {Li}, {Yang}, {Zhang},
  {Zhu}, {Zhang}, {Lu}, {Wang}, {Dai}, {Lynch}, {Yao}, {Jiang}, {Niu}, {Zhou},
  {Xu}, {Miao}, {Niu}, {Meng}, {Qian}, {Tsai}, {Wang}, {Xue}, {Yue}, {Yuan},
  {Zhang}, \& {Zhang}}]{2022Sci...375.1266F}
{Feng}, Y., {Li}, D., {Yang}, Y.-P., {et~al.} 2022{\natexlab{b}}, Science, 375,
  1266, \dodoi{10.1126/science.abl7759}

\bibitem[{{Feng} {et~al.}(2022{\natexlab{c}}){Feng}, {Zhang}, {Li}, {Wang},
  {Tsai}, {Wang}, {Niu}, {Zhang}, {Niu}, {Miao}, {Yuan}, {Zhu}, {Zhang}, {Xue},
  {Yao}, {Yang}, {Cao}, {Chen}, {Zhu}, \& {Zhang}}]{2022ATel15723....1F}
{Feng}, Y., {Zhang}, Y., {Li}, D., {et~al.} 2022{\natexlab{c}}, The
  Astronomer's Telegram, 15723, 1

\bibitem[{{Herrmann}(2022)}]{2022ATel15691....1H}
{Herrmann}, W. 2022, The Astronomer's Telegram, 15691, 1

\bibitem[{{Hilmarsson} {et~al.}(2021{\natexlab{a}}){Hilmarsson}, {Spitler},
  {Main}, \& {Li}}]{2021MNRAS.508.5354H}
{Hilmarsson}, G.~H., {Spitler}, L.~G., {Main}, R.~A., \& {Li}, D.~Z.
  2021{\natexlab{a}}, \mnras, 508, 5354, \dodoi{10.1093/mnras/stab2936}

\bibitem[{{Hilmarsson} {et~al.}(2021{\natexlab{b}}){Hilmarsson}, {Michilli},
  {Spitler}, {Wharton}, {Demorest}, {Desvignes}, {Gourdji}, {Hackstein},
  {Hessels}, {Nimmo}, {Seymour}, {Kramer}, \& {Mckinven}}]{2021ApJ...908L..10H}
{Hilmarsson}, G.~H., {Michilli}, D., {Spitler}, L.~G., {et~al.}
  2021{\natexlab{b}}, \apjl, 908, L10, \dodoi{10.3847/2041-8213/abdec0}

\bibitem[{{Hutschenreuter} {et~al.}(2022){Hutschenreuter}, {Anderson}, {Betti},
  {Bower}, {Brown}, {Br{\"u}ggen}, {Carretti}, {Clarke}, {Clegg}, {Costa},
  {Croft}, {Van Eck}, {Gaensler}, {de Gasperin}, {Haverkorn}, {Heald}, {Hull},
  {Inoue}, {Johnston-Hollitt}, {Kaczmarek}, {Law}, {Ma}, {MacMahon}, {Mao},
  {Riseley}, {Roy}, {Shanahan}, {Shimwell}, {Stil}, {Sobey}, {O'Sullivan},
  {Tasse}, {Vacca}, {Vernstrom}, {Williams}, {Wright}, \&
  {En{\ss}lin}}]{2022A&A...657A..43H}
{Hutschenreuter}, S., {Anderson}, C.~S., {Betti}, S., {et~al.} 2022, \aap, 657,
  A43, \dodoi{10.1051/0004-6361/202140486}

\bibitem[{{Jahns} {et~al.}(2023){Jahns}, {Spitler}, {Nimmo}, {Hewitt},
  {Snelders}, {Seymour}, {Hessels}, {Gourdji}, {Michilli}, \&
  {Hilmarsson}}]{2023MNRAS.519..666J}
{Jahns}, J.~N., {Spitler}, L.~G., {Nimmo}, K., {et~al.} 2023, \mnras, 519, 666,
  \dodoi{10.1093/mnras/stac3446}

\bibitem[{{Jiang} {et~al.}(2020){Jiang}, {Tang}, {Hou}, {Liu}, {Kr{\v{c}}o},
  {Qian}, {Sun}, {Ching}, {Liu}, {Duan}, {Yue}, {Gan}, {Yao}, {Li}, {Pan},
  {Yu}, {Liu}, {Li}, {Peng}, {Yan}, \& {FAST
  Collaboration}}]{2020RAA....20...64J}
{Jiang}, P., {Tang}, N.-Y., {Hou}, L.-G., {et~al.} 2020, Research in Astronomy
  and Astrophysics, 20, 064, \dodoi{10.1088/1674-4527/20/5/64}

\bibitem[{{Keating} \& {Pen}(2020)}]{2020MNRAS.496L.106K}
{Keating}, L.~C., \& {Pen}, U.-L. 2020, \mnras, 496, L106,
  \dodoi{10.1093/mnrasl/slaa095}

\bibitem[{{Kirsten} {et~al.}(2022{\natexlab{a}}){Kirsten}, {Hessels}, {Hewitt},
  {Ould-Boukattine}, {Snelders}, {Gopinath}, {Nimmo}, {Karuppusamy},
  {Herrmann}, {Yang}, {Gawronski}, {Blaauw}, {Buttaccio}, {Maccaferri}, {Bach},
  {Feiler}, {Bray}, {Williams}, {Wrigley}, {Marcote}, {Keimpema}, {Paragi},
  {Burgay}, {Corongiu}, {Giroletti}, {Kramer}, {Pilia}, {Spitler}, {Surcis},
  {Trudu}, {Yuan}, {Wang}, \& {Bezrukovs}}]{2022ATel15727....1K}
{Kirsten}, F., {Hessels}, J.~W.~T., {Hewitt}, D.~M., {et~al.}
  2022{\natexlab{a}}, The Astronomer's Telegram, 15727, 1

\bibitem[{{Kirsten} {et~al.}(2022{\natexlab{b}}){Kirsten}, {Marcote}, {Nimmo},
  {Hessels}, {Bhardwaj}, {Tendulkar}, {Keimpema}, {Yang}, {Snelders}, {Scholz},
  {Pearlman}, {Law}, {Peters}, {Giroletti}, {Paragi}, {Bassa}, {Hewitt},
  {Bach}, {Bezrukovs}, {Burgay}, {Buttaccio}, {Conway}, {Corongiu}, {Feiler},
  {Forss{\'e}n}, {Gawro{\'n}ski}, {Karuppusamy}, {Kharinov}, {Lindqvist},
  {Maccaferri}, {Melnikov}, {Ould-Boukattine}, {Possenti}, {Surcis}, {Wang},
  {Yuan}, {Aggarwal}, {Anna-Thomas}, {Bower}, {Blaauw}, {Burke-Spolaor},
  {Cassanelli}, {Clarke}, {Fonseca}, {Gaensler}, {Gopinath}, {Kaspi}, {Kassim},
  {Lazio}, {Leung}, {Li}, {Lin}, {Masui}, {Mckinven}, {Michilli}, {Mikhailov},
  {Ng}, {Orbidans}, {Pen}, {Petroff}, {Rahman}, {Ransom}, {Shin}, {Smith},
  {Stairs}, \& {Vlemmings}}]{2022Natur.602..585K}
{Kirsten}, F., {Marcote}, B., {Nimmo}, K., {et~al.} 2022{\natexlab{b}}, \nat,
  602, 585, \dodoi{10.1038/s41586-021-04354-w}

\bibitem[{{Kumar} {et~al.}(2023){Kumar}, {Luo}, {Price}, {Shannon}, {Deller},
  {Bhandari}, {Feng}, {Flynn}, {Jiang}, {Uttarkar}, {Wang}, \&
  {Zhang}}]{Kumar+23arXiv}
{Kumar}, P., {Luo}, R., {Price}, D.~C., {et~al.} 2023, arXiv e-prints,
  arXiv:2304.01763, \dodoi{10.48550/arXiv.2304.01763}

\bibitem[{{Li} {et~al.}(2021){Li}, {Wang}, {Zhu}, {Zhang}, {Zhang}, {Duan},
  {Zhang}, {Feng}, {Tang}, {Chatterjee}, {Cordes}, {Cruces}, {Dai}, {Gajjar},
  {Hobbs}, {Jin}, {Kramer}, {Lorimer}, {Miao}, {Niu}, {Niu}, {Pan}, {Qian},
  {Spitler}, {Werthimer}, {Zhang}, {Wang}, {Xie}, {Yue}, {Zhang}, {Zhi}, \&
  {Zhu}}]{2021Natur.598..267L}
{Li}, D., {Wang}, P., {Zhu}, W.~W., {et~al.} 2021, \nat, 598, 267,
  \dodoi{10.1038/s41586-021-03878-5}

\bibitem[{{Lorimer} {et~al.}(2007){Lorimer}, {Bailes}, {McLaughlin},
  {Narkevic}, \& {Crawford}}]{2007Sci...318..777L}
{Lorimer}, D.~R., {Bailes}, M., {McLaughlin}, M.~A., {Narkevic}, D.~J., \&
  {Crawford}, F. 2007, Science, 318, 777, \dodoi{10.1126/science.1147532}

\bibitem[{{Lu} {et~al.}(2020){Lu}, {Kumar}, \& {Zhang}}]{Lu2020}
{Lu}, W., {Kumar}, P., \& {Zhang}, B. 2020, \mnras, 498, 1397,
  \dodoi{10.1093/mnras/staa2450}

\bibitem[{{Luo} {et~al.}(2018){Luo}, {Lee}, {Lorimer}, \& {Zhang}}]{luo2018}
{Luo}, R., {Lee}, K., {Lorimer}, D.~R., \& {Zhang}, B. 2018, \mnras, 481, 2320,
  \dodoi{10.1093/mnras/sty2364}

\bibitem[{{Luo} {et~al.}(2020){Luo}, {Men}, {Lee}, {Wang}, {Lorimer}, \&
  {Zhang}}]{luo2020b}
{Luo}, R., {Men}, Y., {Lee}, K., {et~al.} 2020, \mnras, 494, 665,
  \dodoi{10.1093/mnras/staa704}

\bibitem[{{McKinven} \& {Chime/Frb Collaboration}(2022)}]{2022ATel15679....1M}
{McKinven}, R., \& {Chime/Frb Collaboration}. 2022, The Astronomer's Telegram,
  15679, 1

\bibitem[{{Mckinven} {et~al.}(2022){Mckinven}, {Gaensler}, {Michilli}, {Masui},
  {Kaspi}, {Bhardwaj}, {Cassanelli}, {Chawla}, {Dong}, {Fonseca}, {Leung},
  {Li}, {Ng}, {Patel}, {Petroff}, {Pearlman}, {Pleunis}, {Rafiei-Ravandi},
  {Rahman}, {Sand}, {Shin}, {Scholz}, {Stairs}, {Smith}, {Su}, \&
  {Tendulkar}}]{2022arXiv220509221M}
{Mckinven}, R., {Gaensler}, B.~M., {Michilli}, D., {et~al.} 2022, arXiv
  e-prints, arXiv:2205.09221, \dodoi{10.48550/arXiv.2205.09221}

\bibitem[{{Mckinven} {et~al.}(2023){Mckinven}, {Gaensler}, {Michilli}, {Masui},
  {Kaspi}, {Su}, {Bhardwaj}, {Cassanelli}, {Chawla}, {F.}, {Dong}, {Fonseca},
  {Leung}, {Petroff}, {Pleunis}, {Rafiei-Ravandi}, {Stairs}, {Tendulkar}, {Li},
  {Ng}, {Patel}, {Pearlman}, {Rahman}, {Sand}, \& {Shin}}]{2023arXiv230208386M}
---. 2023, arXiv e-prints, arXiv:2302.08386, \dodoi{10.48550/arXiv.2302.08386}

\bibitem[{{Michilli} {et~al.}(2018){Michilli}, {Seymour}, {Hessels}, {Spitler},
  {Gajjar}, {Archibald}, {Bower}, {Chatterjee}, {Cordes}, {Gourdji}, {Heald},
  {Kaspi}, {Law}, {Sobey}, {Adams}, {Bassa}, {Bogdanov}, {Brinkman},
  {Demorest}, {Fernandez}, {Hellbourg}, {Lazio}, {Lynch}, {Maddox}, {Marcote},
  {McLaughlin}, {Paragi}, {Ransom}, {Scholz}, {Siemion}, {Tendulkar}, {van
  Rooy}, {Wharton}, \& {Whitlow}}]{2018Natur.553..182M}
{Michilli}, D., {Seymour}, A., {Hessels}, J.~W.~T., {et~al.} 2018, \nat, 553,
  182, \dodoi{10.1038/nature25149}

\bibitem[{{Nimmo} {et~al.}(2023){Nimmo}, {Hessels}, {Snelders}, {Karuppusamy},
  {Hewitt}, {Kirsten}, {Marcote}, {Bach}, {Bansod}, {Barr}, {Behrend},
  {Bezrukovs}, {Buttaccio}, {Feiler}, {Gawro{\'n}ski}, {Lindqvist}, {Orbidans},
  {Puchalska}, {Wang}, {Winchen}, {Wolak}, {Wu}, \&
  {Yuan}}]{2023MNRAS.520.2281N}
{Nimmo}, K., {Hessels}, J.~W.~T., {Snelders}, M.~P., {et~al.} 2023, \mnras,
  520, 2281, \dodoi{10.1093/mnras/stad269}

\bibitem[{{Niu} {et~al.}(2022){Niu}, {Aggarwal}, {Li}, {Zhang}, {Chatterjee},
  {Tsai}, {Yu}, {Law}, {Burke-Spolaor}, {Cordes}, {Zhang}, {Ocker}, {Yao},
  {Wang}, {Feng}, {Niino}, {Bochenek}, {Cruces}, {Connor}, {Jiang}, {Dai},
  {Luo}, {Li}, {Miao}, {Niu}, {Anna-Thomas}, {Sydnor}, {Stern}, {Wang}, {Yuan},
  {Yue}, {Zhou}, {Yan}, {Zhu}, \& {Zhang}}]{2022Natur.606..873N}
{Niu}, C.~H., {Aggarwal}, K., {Li}, D., {et~al.} 2022, \nat, 606, 873,
  \dodoi{10.1038/s41586-022-04755-5}

\bibitem[{{Ould-Boukattine} {et~al.}(2022){Ould-Boukattine}, {Herrmann},
  {Gawronski}, {Gopinath}, {Hessels}, {Keane}, {Blaauw}, {Sluman}, {Mulder},
  {McKenna}, {Snelders}, {Kirsten}, \& {Nimmo}}]{2022ATel15817....1O}
{Ould-Boukattine}, O.~S., {Herrmann}, W., {Gawronski}, M., {et~al.} 2022, The
  Astronomer's Telegram, 15817, 1

\bibitem[{{Pelliciari} {et~al.}(2022){Pelliciari}, {Bernardi}, {Pilia},
  {Naldi}, {Bianchi}, {Magro}, {Pupillo}, {Setti}, \&
  {Trudu}}]{2022ATel15695....1P}
{Pelliciari}, D., {Bernardi}, G., {Pilia}, M., {et~al.} 2022, The Astronomer's
  Telegram, 15695, 1

\bibitem[{{Perera} {et~al.}(2022){Perera}, {Perillat}, {Fernandez},
  {Manoharan}, {Roshi}, {Salter}, {Smith}, {Vaddi}, \&
  {McGilvray}}]{2022ATel15734....1P}
{Perera}, B., {Perillat}, P., {Fernandez}, F., {et~al.} 2022, The Astronomer's
  Telegram, 15734, 1

\bibitem[{{Planck Collaboration} {et~al.}(2016){Planck Collaboration}, {Ade},
  {Aghanim}, {Arnaud}, {Ashdown}, {Aumont}, {Baccigalupi}, {Banday},
  {Barreiro}, {Bartlett}, \& et~al.}]{2016A&A...594A..13P}
{Planck Collaboration}, {Ade}, P.~A.~R., {Aghanim}, N., {et~al.} 2016, \aap,
  594, A13, \dodoi{10.1051/0004-6361/201525830}

\bibitem[{{Qu} \& {Zhang}(2023)}]{2023arXiv230209697Q}
{Qu}, Y., \& {Zhang}, B. 2023, \mnras, \dodoi{10.1093/mnras/stad1072}

\bibitem[{{Qu} {et~al.}(2023){Qu}, {Zhang}, \& {Kumar}}]{2023MNRAS.518...66Q}
{Qu}, Y., {Zhang}, B., \& {Kumar}, P. 2023, \mnras, 518, 66,
  \dodoi{10.1093/mnras/stac3111}

\bibitem[{{Rajwade} {et~al.}(2022){Rajwade}, {Wharton}, {Majid}, {Mickaliger},
  {Stappers}, {Breton}, {Lyne}, {Keith}, {Naudet}, {Pearlman}, {Prince},
  {Walker}, \& {Weltevrede}}]{2022ATel15791....1R}
{Rajwade}, K., {Wharton}, R., {Majid}, W., {et~al.} 2022, The Astronomer's
  Telegram, 15791, 1

\bibitem[{{Rajwade} {et~al.}(2020){Rajwade}, {Mickaliger}, {Stappers},
  {Morello}, {Agarwal}, {Bassa}, {Breton}, {Caleb}, {Karastergiou}, {Keane}, \&
  {Lorimer}}]{2020MNRAS.495.3551R}
{Rajwade}, K.~M., {Mickaliger}, M.~B., {Stappers}, B.~W., {et~al.} 2020,
  \mnras, 495, 3551, \dodoi{10.1093/mnras/staa1237}

\bibitem[{{Ravi}(2022)}]{2022ATel15693....1R}
{Ravi}, V. 2022, The Astronomer's Telegram, 15693, 1

\bibitem[{{Ravi} {et~al.}(2022){Ravi}, {Catha}, {Chen}, {Connor}, {Faber},
  {Lamb}, {Hallinan}, {Harnach}, {Hellbourg}, {Hobbs}, {Hodge}, {Hodges},
  {Law}, {Rasmussen}, {Sharma}, {Sherman}, {Shi}, {Simard}, {Squillace},
  {Weinreb}, {Woody}, {Yadlapalli}, {Ahumada}, {Dong}, {Fremling}, {Huang},
  {Karambelkar}, \& {Miller}}]{2022arXiv221109049R}
{Ravi}, V., {Catha}, M., {Chen}, G., {et~al.} 2022, arXiv e-prints,
  arXiv:2211.09049.
\newblock \doarXiv{2211.09049}

\bibitem[{{Sheikh} {et~al.}(2022){Sheikh}, {Farah}, {Pollak}, {Siemion},
  {Cruz}, {Schoultz}, {Hickish}, {Premnath}, {Maddalena}, {DeBoer}, {Gajjar},
  {Donnachie}, {Singh}, {Davis}, {Snodgrass}, \& {Karn}}]{2022ATel15735....1S}
{Sheikh}, S., {Farah}, W., {Pollak}, A.~W., {et~al.} 2022, The Astronomer's
  Telegram, 15735, 1

\bibitem[{{Wang} {et~al.}(2022{\natexlab{a}}){Wang}, {Zhang}, {Dai}, \&
  {Cheng}}]{2022NatCo..13.4382W}
{Wang}, F.~Y., {Zhang}, G.~Q., {Dai}, Z.~G., \& {Cheng}, K.~S.
  2022{\natexlab{a}}, Nature Communications, 13, 4382,
  \dodoi{10.1038/s41467-022-31923-y}

\bibitem[{{Wang} {et~al.}(2022{\natexlab{b}}){Wang}, {Jiang}, {Lee}, {Xu}, \&
  {Zhang}}]{2022MNRAS.517.5080W}
{Wang}, W.-Y., {Jiang}, J.-C., {Lee}, K., {Xu}, R., \& {Zhang}, B.
  2022{\natexlab{b}}, \mnras, 517, 5080, \dodoi{10.1093/mnras/stac3070}

\bibitem[{{Wang} {et~al.}(2022{\natexlab{c}}){Wang}, {Yang}, {Niu}, {Xu}, \&
  {Zhang}}]{2022ApJ...927..105W}
{Wang}, W.-Y., {Yang}, Y.-P., {Niu}, C.-H., {Xu}, R., \& {Zhang}, B.
  2022{\natexlab{c}}, \apj, 927, 105, \dodoi{10.3847/1538-4357/ac4097}

\bibitem[{{Xu} {et~al.}(2022){Xu}, {Niu}, {Chen}, {Lee}, {Zhu}, {Dong},
  {Zhang}, {Jiang}, {Wang}, {Xu}, {Zhang}, {Fu}, {Filippenko}, {Peng}, {Zhou},
  {Zhang}, {Wang}, {Feng}, {Li}, {Brink}, {Li}, {Lu}, {Yang}, {Caballero},
  {Cai}, {Chen}, {Dai}, {Djorgovski}, {Esamdin}, {Gan}, {Guhathakurta}, {Han},
  {Hao}, {Huang}, {Jiang}, {Li}, {Li}, {Li}, {Li}, {Li}, {Liu}, {Luo}, {Men},
  {Niu}, {Peng}, {Qian}, {Song}, {Stern}, {Stockton}, {Sun}, {Wang}, {Wang},
  {Wang}, {Wang}, {Wu}, {Xiao}, {Xiong}, {Xu}, {Xu}, {Yang}, {Yang}, {Yao},
  {Yi}, {Yue}, {Yu}, {Yu}, {Yuan}, {Zhang}, {Zhang}, {Zhang}, {Zhao}, {Zheng},
  {Zhu}, \& {Zou}}]{2022Natur.609..685X}
{Xu}, H., {Niu}, J.~R., {Chen}, P., {et~al.} 2022, \nat, 609, 685,
  \dodoi{10.1038/s41586-022-05071-8}

\bibitem[{{Yang} {et~al.}(2023){Yang}, {Xu}, \& {Zhang}}]{2023MNRAS.520.2039Y}
{Yang}, Y.-P., {Xu}, S., \& {Zhang}, B. 2023, \mnras, 520, 2039,
  \dodoi{10.1093/mnras/stad168}

\bibitem[{{Yang} \& {Zhang}(2018)}]{2018ApJ...868...31Y}
{Yang}, Y.-P., \& {Zhang}, B. 2018, \apj, 868, 31,
  \dodoi{10.3847/1538-4357/aae685}

\bibitem[{{Yao} {et~al.}(2017){Yao}, {Manchester}, \&
  {Wang}}]{2017ApJ...835...29Y}
{Yao}, J.~M., {Manchester}, R.~N., \& {Wang}, N. 2017, \apj, 835, 29,
  \dodoi{10.3847/1538-4357/835/1/29}

\bibitem[{{Yu} {et~al.}(2022){Yu}, {Deng}, {Niu}, {Li}, {Sun}, {Wang}, {Wang},
  {Wu}, \& {Chen}}]{2022ATel15758....1Y}
{Yu}, Z., {Deng}, F., {Niu}, C., {et~al.} 2022, The Astronomer's Telegram,
  15758, 1

\bibitem[{{Zhang}(2022{\natexlab{a}})}]{zhang22}
{Zhang}, B. 2022{\natexlab{a}}, arXiv e-prints, arXiv:2212.03972,
  \dodoi{10.48550/arXiv.2212.03972}

\bibitem[{{Zhang}(2022{\natexlab{b}})}]{2022ApJ...925...53Z}
---. 2022{\natexlab{b}}, \apj, 925, 53, \dodoi{10.3847/1538-4357/ac3979}

\bibitem[{{Zhang} {et~al.}(2021){Zhang}, {Zhang}, {Li}, \&
  {Lorimer}}]{zhangrc2021}
{Zhang}, R.~C., {Zhang}, B., {Li}, Y., \& {Lorimer}, D.~R. 2021, \mnras, 501,
  157, \dodoi{10.1093/mnras/staa3537}

\bibitem[{{Zhang} {et~al.}(2022{\natexlab{a}}){Zhang}, {Niu}, {Feng}, {Zhu},
  {Zhang}, {Di Li}, {Zhu}, {Cao}, {Tsai}, {Li}, {Yu}, {Xu}, {Zhang}, {Lee},
  {Niu}, {Zhou}, {Han}, {Qian}, {Wang}, {Yue}, \& {Yang}}]{2022ATel15733....1Z}
{Zhang}, Y., {Niu}, J., {Feng}, Y., {et~al.} 2022{\natexlab{a}}, The
  Astronomer's Telegram, 15733, 1

\bibitem[{{Zhang} {et~al.}(2018){Zhang}, {Gajjar}, {Foster}, {Siemion},
  {Cordes}, {Law}, \& {Wang}}]{2018ApJ...866..149Z}
{Zhang}, Y.~G., {Gajjar}, V., {Foster}, G., {et~al.} 2018, \apj, 866, 149,
  \dodoi{10.3847/1538-4357/aadf31}

\bibitem[{{Zhang} {et~al.}(2022{\natexlab{b}}){Zhang}, {Wang}, {Feng}, {Zhang},
  {Li}, {Tsai}, {Niu}, {Luo}, {Yao}, {Zhu}, {Han}, {Lee}, {Zhou}, {Niu},
  {Jiang}, {Wang}, {Zhang}, {Xu}, {Wang}, \& {Xu}}]{2022RAA....22l4002Z}
{Zhang}, Y.-K., {Wang}, P., {Feng}, Y., {et~al.} 2022{\natexlab{b}}, Research
  in Astronomy and Astrophysics, 22, 124002, \dodoi{10.1088/1674-4527/ac98f7}

\bibitem[{{Zhao} {et~al.}(2023){Zhao}, {Zhang}, {Wang}, \&
  {Dai}}]{2023ApJ...942..102Z}
{Zhao}, Z.~Y., {Zhang}, G.~Q., {Wang}, F.~Y., \& {Dai}, Z.~G. 2023, \apj, 942,
  102, \dodoi{10.3847/1538-4357/aca66b}

\bibitem[{{Zhou} {et~al.}(2022){Zhou}, {Han}, {Zhang}, {Lee}, {Zhu}, {Li},
  {Jing}, {Wang}, {Zhang}, {Jiang}, {Niu}, {Luo}, {Xu}, {Zhang}, {Wang}, {Xu},
  {Wang}, {Yang}, \& {Feng}}]{2022RAA....22l4001Z}
{Zhou}, D.~J., {Han}, J.~L., {Zhang}, B., {et~al.} 2022, Research in Astronomy
  and Astrophysics, 22, 124001, \dodoi{10.1088/1674-4527/ac98f8}

\end{thebibliography}

\end{document}